\documentclass[12pt,onecolumn,aps,pra,floatfix,superscriptaddress,preprint, showpacs,groupedaddress]{revtex4-1}

\usepackage{graphicx}
\usepackage{graphics}
\usepackage{amssymb}
\usepackage{amsmath}
\usepackage{epsfig}
\usepackage{latexsym}
\usepackage{color}
\usepackage{rotating}
\usepackage{subfigure}
\usepackage{float}
\usepackage[breaklinks=true]{hyperref}

\hypersetup{
  colorlinks   = true, 
  urlcolor     = blue, 
  linkcolor    = blue, 
  citecolor   =  red 
}

\usepackage{soul} 

\begin{document}

\title{Quantum and classical control of single photon states via a mechanical resonator}

\author{Sahar Basiri-Esfahani$^1$\footnote{s.basiri@uq.edu.au}, Casey R. Myers$^1$, {Joshua Combes$^{1,2}$\footnote{Current adresses: Institute for Quantum Computing, Department of Applied Mathematics, University of Waterloo, Waterloo, ON, Canada; Perimeter Institute for Theoretical Physics, 31 Caroline St. N, Waterloo, Ontario, Canada N2L 2Y5.}} and G. J. Milburn$^1$}
\affiliation{$^1$ARC Centre for Engineered Quantum Systems, School of Mathematics and Physics, The University of Queensland, St Lucia, QLD 4072, Australia}
\affiliation{$^2$Center for quantum information and control, University of New Mexico, Albuquerque, New Mexico 87131-0001, USA}

\begin{abstract}
Optomechanical systems typically use light to control the quantum state of a mechanical resonator. In this paper, we propose a scheme for controlling the quantum state of light using the mechanical degree of freedom as a controlled beam splitter. Preparing the mechanical resonator in non-classical states enables an optomechanical Stern-Gerlach interferometer. When the mechanical resonator has a small coherent amplitude it acts as a quantum control, entangling the optical and mechanical degrees of freedom. As the coherent amplitude of the resonator increases, we recover single photon and two-photon interference via a classically controlled beam splitter. The visibility of the two-photon interference is particularly sensitive to coherent excitations in the mechanical resonator and this could form the basis of an optically transduced weak-force sensor.
\end{abstract}

\maketitle

\section{Introduction}

Over the last decade, the new domain of engineered quantum systems has attracted considerable research interest~\cite{warren1993coherent,rabitz2000whither,palomaki2013entangling}, the objective of which is to control the quantum world. Quantum engineered systems are a promising emergent technology, with many important applications such as enhanced sensing and metrology~\cite{Kolkowitz,Gavartin,Maiwald,BasiriOpx}, fundamental tests of quantum mechanics~\cite{jeong2014coarsening} and quantum information and quantum computation~\cite{bennett2000quantum,knill2001scheme}. Quantum control involves manipulating the evolution of a quantum system to steer the system to a desired state. This is particularly important in quantum information and computation in which quantum states of a qubit, such as a single photon, must be prepared, controlled and measured.

In quantum optomechanics~\cite{BowenMilburn}, optical fields are used to both control the state of the mechanical system and to read it out. In the simplest case, sideband cooling is used to prepare the mechanical resonator in its ground state. From there, various schemes have been devised to steer the mechanical resonator into non-classical states such as squeezed states~\cite{Wollman,Pirkkalainen,Lecocq}, phonon number eigenstates~\cite{Galland,Vanner,Borkje,NJP-single-photon} and cat states~\cite{Akram,Setkaski} by suitable optical control. Given the ability to prepare a mechanical resonator in a non-classical state, we can then consider the possibility of using it to control the quantum state of the light. It is therefore interesting to investigate the control of single photon states and photon-photon interactions mediated by a mechanical resonator prepared in a non-classical state with possible applications in photonic quantum information processing~\cite{Agarwal,Stannigel,Komar}. In our protocol, the mechanical state can be prepared using a Raman process very much like that used for atomic Raman memories. 

Single photon states~\cite{Milburn-singlePhot} are predominantly employed as information carriers in quantum communication~\cite{Scarani} and quantum information processing~\cite{langford2011efficient}. This has recently motivated the development of single photonic technologies such as the methods to generate, control, process and measure single photons~\cite{Collins,Buckley,Riedinger,Comandar,Ralph2006}. At the heart of single photon experiments is the interference of two single photons at a beam splitter, a uniquely quantum feature first demonstrated by Hong, Ou and Mandel and known as HOM interference~\cite{HOM}. It lies at the core of the power of linear optical quantum information processing~\cite{knill2001scheme}, demonstrated most recently in boson sampling experiments~\cite{boson-sampling1,boson-sampling2}. This effect is observed when two indistinguishable single photons enter a 50/50 beam splitter where the photons bunch together and both are detected at one output port as a result of quantum interference. As HOM interference is a purely quantum mechanical manifestation of single photon states interacting at a beam splitter, controlling the beam splitter interaction would lead to quantum or classical control of the single photon states.

Langford et al.~\cite{langford2011efficient} have introduced a controlled beam splitter interaction to coherently control the conversion of photons between three optical modes as a route to generate and process complex, multi-quanta states for photonic quantum information applications. In this paper, we consider this model in an optomechanical scheme to take advantage of the intrinsic optomechanical nonlinearities for coherent photon conversion and controlling the non-classical states of light. Our model considers a quantum controlled beam splitter in which a mechanical degree of freedom controls a beam splitter interaction between two optical modes. In particular, we show that when the mechanical resonator is prepared in a phonon number state it acts as a quantum controller, entangling the optical and mechanical degrees of freedom; a kind of optomechanical Stern-Gerlach interferometer. When the coherent excitation of the mechanical resonator is increased, the controller acts as a classical parameter, resulting in a classical beam splitter interaction. The experimental signature of the transition from quantum control to classical control is the visibility achievable in a Mach-Zehnder (MZ) or Hong-Ou-Mandel (HOM) interferometer for one and two photon inputs, respectively. The results we present demonstrate how an underlying quantum control can be configured as a purely classical control provided the residual entanglement between the controller and the system can be made arbitrarily small~\cite{Milburn-RS}.

The implementation of single photon optomechanics entails a strong coupling between the mechanical and optical modes and is not yet experimentally achieved. However, several groups have put some efforts towards this~\cite{Riedinger,Yuan}. For example, photonic crystal (PhC) resonators make it possible to get both localized optical and mechanical modes at the same time. This increases the optomechanical coupling strength between the optical and vibrational mode~\cite{Painter,Chang}. Therefore, PhCs are promising candidates for engineering strong single photon optomechanical coupling by providing high-Q nano-cavities. This achievement, together with improvements to single photon sources~\cite{stock2011high,buckley2012engineered} and the technology to couple single photons into PhC cavities~\cite{Collins}, offers a platform for novel proposals using single photon optomechanics. 

Our proposal is based on a double cavity optomechanical system wherein a mechanical resonator modulates the coherent coupling of the two optical cavities. Each cavity has a single input-output channel. This system offers a strong intrinsic nonlinearity which cannot be achieved using standard linear optical components. This nonlinearity can be used to implement mechanically assisted coherent photon conversion between the two different optical modes and create an effective photon-photon interaction. An example of this kind of system has been developed by the Painter group~\cite{Chang}. Another example is based on a single bulk flexural mode driven by the opposing radiation pressure forces of two optical cavity modes. If the cavity modes are coupled, transformation to normal modes leads to a model in which the normal mode coupling is modulated by the mechanical displacement~\cite{Marquardt2012}.

The protocol we describe is based on the ability to prepare the mechanical resonator in either a Fock state or a coherent state by transferring the desired state from the optics to the mechanics using one of the optical modes. In essence, this preparation step is using the mechanics as a quantum memory and parallels atomic Raman memory schemes~\cite{Nunn2008}. In this way, we can prepare the mechanics in a single phonon state or a coherent state with varying amplitude. In the second stage of the protocol, we investigate how the prepared mechanical state controls the beam splitter interaction between two optical modes prepared in single photon states. We show that if the mechanical resonator is described classically, this interaction implements a controllable beam splitter interaction between the input and output modes of the optical cavities. As the coherent amplitude of the mechanical resonator is reduced, the photons become entangled with the mechanical resonator and this is reflected in a decrease in the visibility of a MZ or HOM interferometer. Optical interferometry is thus a probe of the entanglement between a quantum controller and the target system.

The paper is organised as follows. In section \ref{CPC}, we introduce a model for a mechanically controlled beam splitter and, using a simple unitary model  encoding a qubit or a qutrit into the optical degrees of freedom, we investigate how the state of the controller can be varied to enable quantum or classical control of the optical system with the transition between these extremes determined by the degree of entanglement  between the optical and mechanical subsystems. A simple interpretation is given in terms of `which-path' information stored in the controller. In section \ref{single-photon}, we define continuous mode single photon states of the field. In section \ref{OM_model}, we generalize the simple single frequency analysis from section \ref{CPC} to a multimode input-output analysis. In section \ref{mech-intfr}, we show how the degree of quantum control can be determined using optical interferometry in which the mechanical system acts as a controlled beam splitter in place of a conventional beam splitter. The mechanically controlled beam splitter is comprised of two coupled optical cavities with a coupling rate that depends on the mechanical displacement and with each cavity coupled to a single input mode and a single output mode. The visibility of the resulting interferometer is shown to be an experimental signature of the degree of entanglement between the state of the mechanical element and the light. As the mechanical element becomes a classical controller, the entanglement goes to zero and the interferometer visibility goes to maximum value. Section \ref{discussion} discusses the broader implications of our model.

\section{Coherent photon conversion.}
\label{CPC}

In this section, we provide a simple model which qualitatively explains many features that arise in the fully quantum model presented in section \ref{OM_model}. The model consists of  three single (frequency) mode bosons interacting via a cubic interaction Hamiltonian. Two of the bosonic modes describe the optical cavities and the third bosonic mode describes a mechanical system. We show that  by preparing the mechanical system in various states (e.g. a Fock or coherent state) we can control the interaction between the optical modes.

In section \ref{unitary_model}, we present the model and review the concept of coherent photon conversion~\cite{langford2011efficient}. In section \ref{semiclassical}, we explicitly show that if the mechanical object is prepared in a large amplitude coherent state, to a very good approximation, it does not encode any `which-path' information about the optical mode excitations while a perturbative analysis of the residual quantum correlations between the optical and mechanical modes indicates that which-path information is present at higher orders in the coupling strength. Finally, section \ref{Q_control} illustrates that if the mechanics is in a quantum state such as a Fock state or a small amplitude coherent state, the optical and mechanical degrees of freedom become entangled. This entanglement changes the behaviour of the system by varying the extent to which which-path information is present for different initial states of the controller. 

\subsection{Unitary model}\label{unitary_model}
The classical beam splitter interaction between two optical modes $a_1$ and $a_2$ is defined by the unitary operator
\begin{equation}
\label{BS}
U_{{\rm BS}}(\theta)=e^{-i\theta (a_1^\dagger a_2+a_1 a_2^\dagger)},
\end{equation}
under which the optical operators transform as
\begin{eqnarray}
U_{{\rm BS}}^\dagger(\theta) a_1 U_{{\rm BS}}(\theta) & = & \cos(\theta)\ a_1-i\sin(\theta)a_2 ,\\
U_{{\rm BS}}^\dagger(\theta) a_2 U_{{\rm BS}}(\theta) & = & \cos(\theta)a_2-i\sin(\theta)a_1 .
\end{eqnarray}

Langford et al.~\cite{langford2011efficient} introduced the concept of coherent photon conversion based on an ability to coherently control the exchange of photons between two optical modes. The Hamiltonian realized an optical three-wave mixing process in which a strong coherent pump field was used to create an enhanced nonlinearity in the nonlinear medium to convert a single photon into two single excitations in different frequencies. The defining Hamiltonian for the process is
\begin{equation}
 H_{\rm I}= \hbar g(a_1^\dagger a_2  b^\dagger +a_1 a_2^\dagger  b),
 \label{interaction-Hamiltonian}
\end{equation}
where $a_1,a_2$ are the annihilation operators for the bosonic modes we seek to control while $b$ is the annihilation operator of the bosonic controller. In~\cite{langford2011efficient}, the controller was taken to also be an optical mode but in this paper the  controller will represent a mechanical degree of freedom. Our scheme uses the intrinsic nonlinearity of the optomechanical beam splitter and does not need a nonlinear crystal.

The Hamiltonian in equation (\ref{interaction-Hamiltonian}) will, in general, dynamically entangle the optical and mechanical degrees of freedom depending on the initial states used. We will now describe a picture in which this entanglement can be viewed in terms of `which-path' information in a kind of optical Stern-Gerlach device in which optical information is stored in the mechanical element. 

We will assume that the optical modes begin in an eigenstate of the total photon number $N= a_1^\dagger a_1+a_2^\dagger a_2$, while the mechanical element is prepared in an arbitrary state $|\phi\rangle_b$.  It is convenient in this case to use the two mode Schwinger representation of ${\rm SU}(2)$ to write the joint state of the optical modes. Defining the generators of ${\rm SU}(2)$ as 
\begin{eqnarray}
S_z & = & \frac{1}{2}( a_2^\dagger a_2-a_1^\dagger a_1),\\
S_x & = & \frac{1}{2}( a_2^\dagger a_1+a_1^\dagger a_2),\\
S_y & = & -\frac{i}{2}( a_2^\dagger a_1-a_1^\dagger a_2),
\end{eqnarray}
with $S^2 =(N/2+1)N/2$. We can then define the joint eigenstates of $S^2$ and $S_z$ in terms of the tensor product photon number basis for modes $a_k$ as $|s,m\rangle_z= |s-m\rangle_1\otimes|s+m\rangle_2$. 

The interaction Hamiltonian equation (\ref{interaction-Hamiltonian}) can then be written in the form
\begin{equation}
H_{\rm I}=\hbar g(S_+ b+S_- b^\dagger),\label{tavis}
\end{equation}
where $S_+=S_{-}^\dagger= a_2^\dagger a_1$ or equivalently  $S_\pm = S_x \pm i S_y$. If the initial state of the entire system is
\begin{equation}
|\Psi(0)\rangle= |\psi(0)\rangle_{{\rm O}}\otimes|\phi(0)\rangle_b,\ \ \ 
\end{equation}
the total state at time $t>0$ can then be written as 
\begin{equation}
\label{joint-state}
|\Psi(t)\rangle =U(t)|\Psi(0)\rangle=\sum_{m=-s}^s|s,m\rangle_z\otimes|\phi_m(t)\rangle_b,
\end{equation}
 with $ U(t) = e^{-\frac{i}{\hbar}H_{\rm I}t}$ and 
 \begin{equation}
 \label{phi_m(t)}
 |\phi_m(t)\rangle_b= \mbox{}_z\langle s,m|U(t)|\psi(0)\rangle_{{\rm O}}\otimes|\phi(0)\rangle_b.
 \end{equation}
  Equation (\ref{joint-state}) is like a Stern-Gerlach device in which the mechanical controller is keeping track of `which-path' information. We make this interpretation more explicit in section \ref{semiclassical} below. 

Clearly $|\Psi(t)\rangle$ is an entangled state in general. Tracing out the state of the controller, we see that the state of the optical system is 
\begin{equation}
\label{optics}
 \rho_{{\rm O}}(t)=\sum_{m,n=-s}^s R_{n,m}(t)   |s,m\rangle_z\langle s,n|,
 \end{equation}
 where
 \begin{equation}
 \label{coefficients}
 R_{nm}(t) =\mbox{}_b\langle \phi_n(t)|\phi_m(t)\rangle_b.
 \end{equation}
In general, $ \rho_{{\rm O}}(t)$ is a mixed state representing the degree of entanglement between the optical system and the controller; a correlation that encodes `which-path' information if the states $|\phi_m(t)\rangle_b$ are distinguishable.

 \subsection{Classical control}\label{semiclassical}
We now consider the case in which the mechanical element is prepared in a coherent state, $|\beta\rangle_b$, with a large coherent amplitude $\beta$. We can use a canonical transformation $
b\rightarrow \bar{b}+\beta,$ to include this amplitude in the Hamiltonian while the initial mechanical state now becomes the ground state. The new interaction Hamiltonian is
\begin{equation}
\label{displaced-ham}
H_{\rm I}=\hbar \bar{g}(a_1^\dagger a_2 +a_1 a^\dagger_2)+ \hbar g (a_1^\dagger a_2 \bar{b}^\dagger+a_1 a^\dagger_2\bar{b} ).
\end{equation}
where $\bar{g}=g\beta$.  For simplicity, and without loss of generality, we take $\beta$ to be real. 

As a coherent state with large amplitude is a semiclassical state, we expect that as $|\beta|$ becomes large, this model should reduce to the simple unitary model of equation (\ref{BS}). To see this, we fix  $\bar{g}$ to be constant while letting $\beta\rightarrow \infty$. With this scaling we can regard $g=\dfrac{\bar{g}}{\beta} \ll 1$ as a perturbation parameter in the dynamics arising from equation (\ref{displaced-ham}). Terms arising to first and higher order terms in $g$ describe optomechanical entanglement and a departure from the simple unitary coupling of the optical modes described by equation (\ref{BS}). 

We define $\theta=\bar{g}t$ as the effective beam splitter parameter that can be reached by unitary evolution; for example, a $50/50$ beam splitter has $\theta=\pi/4$. Let the initial state of the optics and the mechanics be 
\begin{equation}
|\Psi(0)\rangle = |\psi(0)\rangle_{{\rm O}}\otimes|\beta\rangle_b,
\end{equation}
where $|\psi(0)\rangle_{{\rm O}}=|n\rangle_1\otimes |m\rangle_2$ is the initial state of the mode-1 and mode-2, taken as a product Fock state, and $|\beta\rangle_b$ is a coherent state. In terms of the ${\rm SU}(2)$ operators, $|\psi(0)\rangle_{{\rm O}}$ is an eigenstate of $\hat{S}_z$. Therefore, equation (\ref{phi_m(t)}) can be written as
\begin{equation}
 |\phi_m(t)\rangle_b = D(\beta) \mbox{}_z\langle s,m|\bar U(\theta)|\psi(0)\rangle_{{\rm O}}\otimes|0\rangle_b,
\end{equation} 
where
\begin{equation}
\bar U(\theta)=D^\dagger(\beta)e^{-igt(S_{+}b+S_{-}b^\dagger)}D(\beta)=e^{-2i\theta S_x-igt(S_{+}\bar{b}+S_{-}\bar{b}^\dagger)},
\end{equation}
and $D(\beta)$ is the displacement operator with the property $D^\dagger(\beta)bD(\beta)=\bar{b}+\beta$. To second order in $gt$, we find that 
\begin{eqnarray}
|\phi_m(\theta)\rangle &=& \mbox{}_z\langle s,m|U_{{\rm BS}}|\psi(0)\rangle_{\rm O} |\beta\rangle_b+\theta gt\mbox{}_z\langle s,m|U_{{\rm BS}}S_z|\psi(0)\rangle_{\rm O} D(\beta)|1\rangle_b\nonumber\\
& & +\frac{(\theta gt)^2}{2}\mbox{}_z\langle s,m|U_{{\rm BS}}S^2_z|\psi(0)\rangle_{\rm O}\left (\sqrt{2}D(\beta)|2\rangle_b-|\beta\rangle_b\right)+...,
\end{eqnarray}
where $U_{\rm BS}$ is given by equation (\ref{BS}).
Substituting this into equation (\ref{optics}), we then find that 
 \begin{equation}
 \label{dephasing}
 \rho_{{\rm O}}(\theta)=U_{{\rm BS}}(\theta)\left (\rho_{{\rm O}}(0)-\frac{(\theta gt)^2}{2}[S_z,[S_z,\rho_{{\rm O}}(0)]]+...\right )U^\dagger_{{\rm BS}}(\theta)
 \end{equation}
 with $\rho_{{\rm O}}(0)=|\psi(0)\rangle_{\rm O}\langle \psi(0)|$. The state of the two optical cavities, after fixed interaction time such that $\theta=\bar{g}t= {\rm constant}$, is thus given by a completely positive unital map of the initial state.
We can now see that, up to second order correction to the action of a classical beam splitter, we achieve a completely positive map corresponding to a dephasing channel that might arise, for example, as a weak measurement of $S_z$. This is precisely what one would expect if the mechanics encoded `which-path' information about the optical excitations. The classical control given by the beam splitter interaction $U_{\rm BS}$ is composed with a random relative phase shift between the two optical cavities. This leads to a weak suppression of the off-diagonal coherence terms in the $S_z$ basis. This effect can be interpreted as an effective measurement of the relative occupation number of each cavity mode.  The results of this measurement are hidden in the quantum correlations between the cavity modes and the mechanical degree of freedom and thus reflects residual entanglement between the controller and the optical system. This structure is typical of the way in which residual quantum entanglement in a semiclassical controller affects the ideal classical control transformation~\cite{Milburn-RS}. The dephasing channel described above is {\em not} the only decoherence phenomena that can happen (to second order in $gt$). However, this expression shows the point that once we go beyond zero order, the controller gets entangled with the optical system.

 \subsection{Quantum control}\label{Q_control}
We now consider the opposite limit in which the mechanical element is a quantum controller for the optical states. Quantum control necessarily requires that the controller becomes entangled with the target state for appropriate states of the controller. For example, in a quantum CNOT gate, preparing the controller in a superposition of the two computational basis states with the target in one of the computational basis states, produces a Bell state for the combined system. 
 
As in the previous section, the Schwinger representation allows one to see the kind of quantum control realized in this system. We now write the Hamiltonian given in equation (\ref{interaction-Hamiltonian}) and equation (\ref{tavis}), in terms of the (dimensionless) position and momentum operators for the mechanics
\begin{equation}
H_{\rm I} =\hbar g(S_xX-S_yY),
\end{equation}
where $X=b+b^\dagger,\ Y =-i(b-b^\dagger)$. This Hamiltonian represents two orthogonal rotations of the pseudo-spin controlled by two non-commuting operators in the controller. This is the canonical example of quantum control considered in \cite{HMW-GJM}: it is not possible to represent this kind of control using a measurement and feedback protocol from the controller to the optical subsystem.
 
 To proceed,  we need to fix the initial optical state. For practical reasons it is unlikely that we will be able to prepare states with $N>2$ in the foreseeable future. The case $N=1$  requires a single photon and encodes a qubit in the two optical modes~\cite{NJP-single-photon}. In the next section, we will see that this can be done using a MZ interferometer set-up.  The case $N=2$ can be done by preparing a single photon in each optical mode and corresponds to encoding a qutrit into the optical system. In this case, the initial state of the optical system is $|1\rangle_1\otimes|1\rangle_2\equiv |1,0\rangle_z$ in the Schwinger representation. In the next section, we will see that this suggests a HOM interferometer set-up to investigate quantum to classical control. 
 
 The connection to interferometry can be seen more clearly by writing equation (\ref{joint-state}) for the two cases $N=1,2$ with the initial optical states $|0\rangle_1|1\rangle_2$ and $|1\rangle_1|1\rangle_2$ respectively 
 \begin{eqnarray}
 |\Psi_{1/2}(t)\rangle & = & |1/2, -1/2\rangle_z|\phi_{-1/2}(t)\rangle_b+ |1/2, 1/2\rangle_z|\phi_{1/2}(t)\rangle_b,\\
  |\Psi_{1}(t)\rangle & = & |1, -1\rangle_z|\phi_{-1}(t)\rangle_b+ |1, 0\rangle_z|\phi_{0}(t)\rangle_b+|1, 1\rangle_z|\phi_{1}(t)\rangle_b.
  \end{eqnarray}
These two equations indicate that the mechanical element acts, in general, as quantum controller of a qubit  ($N=1$) or a qutrit ($N=2$).
  
  In the $N=1$ qubit case, it is easier to work in the dressed state basis rather than the tensor product basis as the dressed states are eigenstates of the Hamiltonian. These are defined by $H_{\rm I}|\pm, n\rangle =\pm  \hbar g\sqrt{n}|\pm, n\rangle$, where
  \begin{eqnarray}
  |+,n\rangle & = & \frac{1}{\sqrt{2}}(|1/2, n-1\rangle+|-1/2, n\rangle),\\
  |-,n\rangle & = &  \frac{1}{\sqrt{2}}(|1/2, n-1\rangle-|-1/2, n\rangle),
  \end{eqnarray}
  where $|\pm 1/2, n\rangle = |1/2, \pm1/2\rangle\otimes|n\rangle_b$ with $|n\rangle_b$ a Fock state of the mechanical oscillator.  If the initial state is written in the dressed-state basis as
  \begin{equation}
  |\psi(0)\rangle_{{\rm O}}\otimes|\phi(0)\rangle_b = \sum_n c_n^+ |+,n\rangle +c_n^-|-,n\rangle,
  \end{equation}
where $c^{\pm}_n$ are complex coefficients, the state at a later time is 
   \begin{equation}
   |\Psi(t)\rangle = \sum_n c_n^+ e^{-ig\sqrt{n} t}|+,n\rangle +c_n^-e^{ig\sqrt{n} t} |-,n\rangle,
   \end{equation}
which is clearly a controlled rotation in the dressed state basis. For example, if the optical system is prepared in the state $|1/2,-1/2\rangle$ while the mechanical controller is prepared with a single excitation, $|1\rangle_b$, the state at a later time is 
   \begin{equation}
   \label{state_Qbit1}
   |\Psi(t)\rangle=-i\sin(gt)|1/2,1/2\rangle\otimes |0\rangle_b+\cos(gt)|1/2,-1/2\rangle\otimes|1\rangle_b,
   \end{equation}
an entangled state in general. 
   
   In the case $N=2$, the qutrit case, the corresponding case for the initial state 
   \begin{equation}
    |\Psi(0)\rangle = |1,0\rangle_z\otimes|1\rangle_b,
    \end{equation}
    is 
 \begin{eqnarray}
 \label{state_Qtrit1}
 |\Psi(t)\rangle &=& -i\sqrt{\frac{1}{3}}\sin(\sqrt{6}gt)\vert 1,1\rangle_z\otimes|0\rangle_b+\cos(\sqrt{6}gt)\vert 1,0\rangle_z\otimes|1\rangle_b\nonumber\\&&-i\sqrt{\frac{2}{3}}\sin(\sqrt{6}gt)|1,-1\rangle_z\otimes|2\rangle_b.
 \end{eqnarray}
If we write the state of the optical system in form of equation (\ref{optics}), we have
\begin{equation}
\rho_{\rm O}(t)=R_{1,1}(t)\vert 1,1\rangle_z\langle 1,1\vert+R_{0,0}(t)\vert 1,0\rangle_z\langle 1,0\vert+R_{-1,-1}(t)\vert 1,-1\rangle_z\langle 1,-1\vert,
\end{equation}
where
\begin{equation}
R_{1,1}(t)=\frac{1}{3}\sin^2(\sqrt{6}gt),\hspace{3mm}R_{0,0}(t)=\cos^2(\sqrt{6}gt),\hspace{3mm}R_{-1,-1}(t)=\frac{2}{3}\sin^2(\sqrt{6}gt).
\end{equation}
We see that $R_{n,m}(t)=0$ for $n\neq m$ and there is a complete loss of coherence, due to the entanglement of the optical and mechanical systems,  and  perfect which-path information of the optical system is encoded in the mechanical object. Inspection of equation (\ref{state_Qtrit1}) indicates that this information is stored in the number of mechanical excitations.

If the optical system is prepared in the same initial state in each case $N=1,2$ but the mechanical controller is prepared in a superposition of ground and single excitation, $\frac{1}{\sqrt{2}}(\vert 0\rangle_b+\vert 1\rangle_b)$, the state at time $t$ for the case $N=1$ becomes
\begin{equation}
   \label{state_Qbit2}
   |\Psi(t)\rangle=\frac{1}{\sqrt{2}}(|1/2,-1/2\rangle-i\sin(gt)|1/2,1/2\rangle)\otimes |0\rangle_b+\frac{1}{\sqrt{2}}\cos(gt)|1/2,-1/2\rangle\otimes|1\rangle_b,
\end{equation}
   and for the case $N=2$ we have
\begin{eqnarray}
    \label{state_Qtrit2}
 |\Psi(t)\rangle &&=\Big({\frac{-i}{\sqrt{6}}}\sin(\sqrt{6}gt)|1,1\rangle_z+\frac{1}{\sqrt{2}}\cos(\sqrt{2}gt)\vert1,0\rangle_z)\Big)\otimes|0\rangle_b\nonumber\\
 &&\ +\Big(\frac{1}{\sqrt{2}}\cos(\sqrt{6}gt)|1,0\rangle_z-\frac{i}{\sqrt{2}}\sin(\sqrt{2}gt)\vert1,-1\rangle_z\Big)\otimes|1\rangle_b\nonumber\\&&\ -\frac{i}{\sqrt{9}}\sin(\sqrt{6}gt)|1,-1\rangle_z\otimes|2\rangle_b.
\end{eqnarray}
 \begin{figure}[!htbp]
 \includegraphics[width=11cm]{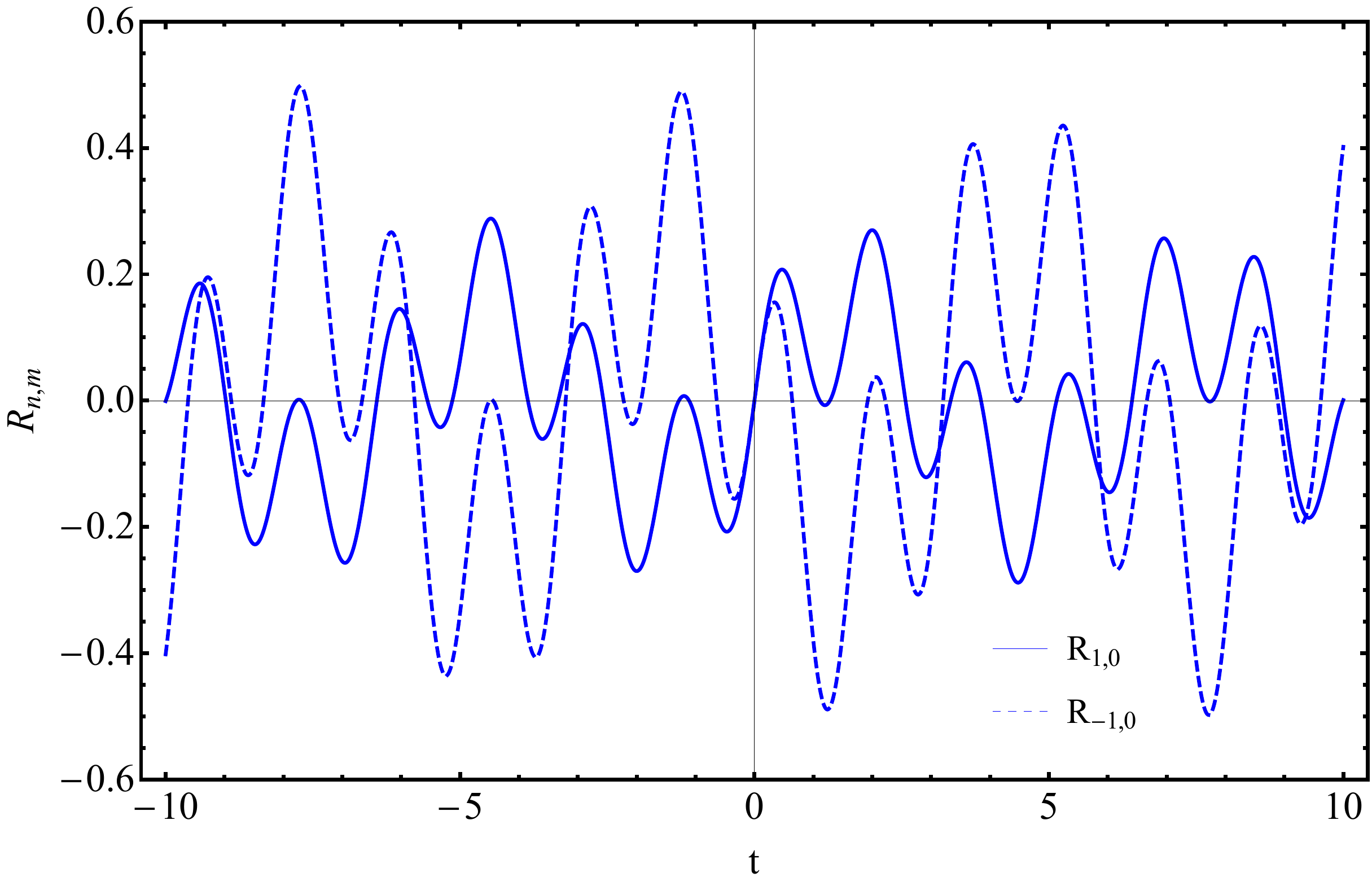}
\caption{The off-diagonal coefficients $R_{n,m}$ versus the interaction time.} 
\label{fig-control}
\end{figure}
Again we take the example for the case $N=2$ to calculate the state of the optical system in form of equation (\ref{optics}) as
\begin{eqnarray}
\rho_{\rm O}(t)=&&R_{1,1}(t)\vert 1,1\rangle_z\langle 1,1\vert+R_{0,0}(t)\vert 1,0\rangle_z\langle 1,0\vert+R_{-1,-1}(t)\vert 1,-1\rangle_z\langle 1,-1\vert\nonumber\\&&+R_{0,1}(t)\vert 1,1\rangle_z\langle 1,0\vert+R_{1,0}(t)\vert 1,0\rangle_z\langle 1,1\vert\nonumber\\&&+R_{0,-1}(t)\vert 1,-1\rangle_z\langle 1,0\vert+R_{-1,0}(t)\vert 1,0\rangle_z\langle 1,-1\vert,
\end{eqnarray}
where
\begin{eqnarray}
&&R_{1,1}(t)=\frac{1}{6}\sin^2(\sqrt{6}gt),\hspace{5mm}R_{0,0}(t)=\frac{1}{2}(\cos^2(\sqrt{6}gt)+\cos^2(\sqrt{6}gt)),\nonumber\\&&R_{-1,-1}(t)=\frac{1}{2}\sin^2(\sqrt{6}gt)+\frac{1}{3}\sin^2(\sqrt{6}gt),\nonumber\\&&R_{0,1}(t)=R^*_{1,0}(t)=\frac{-i}{2\sqrt{3}}\sin(\sqrt{6}gt)\cos(\sqrt{2}gt),\nonumber\\&&R_{0,-1}(t)=R^*_{-1,0}(t)=\frac{-i}{2}\sin(\sqrt{2}gt)\cos(\sqrt{6}gt).
\end{eqnarray}
Figure \ref{fig-control} shows the coefficients $R_{n,m}$ ($n\neq m$) which measure the decoherence. It can be seen that these terms are not equal to zero at the same time thus there is always some coherence in the system which results in the reduction of the which-path information. These examples show that, by varying the state of the controller, the degree of entanglement between the optical system and the controller varies and so does the which-path information which affects the reduced state of the optical system. 

A coherent state with a very small amplitude, $\beta\ll 1$, can also be approximated as an asymmetric superposition of $\vert 0\rangle$ and $\vert 1\rangle$
\begin{equation}
\vert\beta\rangle\simeq \frac{\vert 0\rangle+\beta \vert1\rangle}{\sqrt{1+\vert\beta\vert^2}}.
\end{equation}
Therefore, for a mechanical system prepared in coherent state with small amplitude there is less entanglement between the controller and the optical system than there is for the mechanical number state $|0\rangle_b$ or $|1\rangle_b$, resulting in less decoherence and less which-path information stored in the mechanics.  In section \ref{mech-intfr}, we consider the visibility of the interference pattern, in a MZ interferometer for the qubit case and in a HOM interferometer for the optical qutrit, as a measure of this which-path information.

\section{continuous mode single photon states}
\label{single-photon}
We wish to consider the case in which the field driving each cavity is made up of a sequence of pulses with exactly one photon per pulse. 
The positive frequency components of the input field operator, $a_{\rm in}(t)$ can be written in the frequency domain as 
\begin{equation}
\label{input-noise}
a_{\rm in}(t) =e^{-i\omega_c t}\int_{-\infty}^\infty d\omega\ \tilde{a}_{\rm in}(\omega) e^{-i\omega t},
\end{equation}
where $\omega_c$ is an appropriate carrier frequency. We will work in an interaction picture rotating at the carrier frequency and ignore the oscillatory pre-factor in equation (\ref{input-noise}). However, it should be noted that whenever we use single photon states as an input to an optical cavity we will assume that the carrier frequency is resonant with the cavity. 

We now define a single photon state as a superposition of a single excitation over many frequencies~\cite{milburn2008coherent}
\begin{equation}
|1_{\tilde{\xi}}\rangle=\int_{-\infty}^\infty d\omega\  \tilde{\xi}(\omega)\tilde{a}^\dagger_{\rm in}(\omega)|0\rangle,
\end{equation}
where $\tilde{\xi}(\omega)$ is the spectral density function. The average field amplitude of a single photon state is zero
\begin{equation}
\langle 1_{\tilde{\xi}}|a_{\rm in}(t)|1_{\tilde{\xi}}\rangle =0.
\end{equation}
We can interpret this result as an indication of the random optical phase of a photon number eigenstate. A phase dependent measurement on the single photon state using, for example, homodyne detection would give a null signal on average. Despite this result, it is clear that the single photon state is a pure quantum state and as such contains a great deal of quantum coherence. This is revealed when we look at the intensity of the field rather than the field amplitude. 

The probability per unit time to detect a single photon on an ideal detector is proportional to $n(t)= \langle 1_{\xi}\vert a_{\rm in}^\dagger(t)a_{\rm in}(t)\vert 1_{\xi}\rangle$
and it is easy to show that 
\begin{equation}
\label{intensity}
n(t)=|\xi(t)|^2,
\end{equation}
where 
\begin{equation}
\xi(t)=\int_{-\infty}^\infty d\omega\ e^{-i\omega t}\tilde{\xi}(\omega ).
\end{equation}
The fact that $n(t)$ appears as the modulus square of a single, complex valued function in equation (\ref{intensity}) is a reflection of the underlying purity of the single photon state. In optical terms we would say that the pulse is `transform limited' although we need to bear in mind that this is highly non-classical state with an average field amplitude of zero.

\section{Optomechanical model}\label{OM_model}

The previous discussion, while a good introduction to the central features of the model,  is not realistic from an experimental perspective. Typically, we do not have control of the optical state of the two cavity fields directly, rather we only have control over single photon sources external to the optical cavities.  In this setting, we need to take into account the stochastic nature of the reflection and absorption of photons by the cavities. This situation cannot be described by a {single-mode and} purely Hamiltonian model. Externally driven optical cavities have, of course, long been understood in quantum optics however typically the external fields are coherent or thermal. Here we will introduce new techniques for dealing with the non-stationary input fields that correspond to single photon sources rather than coherent or thermal sources.  

The specific model of the bosonic control field we propose is based on an optomechanical system of two coupled cavity modes with an interaction strength that depends upon a mechanical displacement coordinate, although other models are possible, e.g. a Raman atomic memory. We will model the interaction between the two optical modes and the mechanical resonator in terms of the same third order bosonic Hamiltonian discussed in equation (\ref{interaction-Hamiltonian}). Let $a_1,a_2$ be the annihilation operators for the optical fields in two cavity modes with resonant frequencies $\omega_1,\omega_2$ respectively, while $b$ is the annihilation operator for a mechanical resonator with resonant frequency $\omega_m$. The Hamiltonian for this optomechanical system can be written as~\cite{Chang}
\begin{equation}
H=\hbar\omega_1 a^\dagger_1 a_1+\hbar\omega_2 a_2^\dagger a_2 +\hbar\omega_m b^\dagger b+\hbar g(b+b^\dagger) (a_1^\dagger a_2+a_1 a_2^\dagger).
\end{equation}
We now move to an interaction picture for both optical modes and the mechanical mode. After the rotating wave approximation, the interaction Hamiltonian in the interaction picture including only the resonant terms is 
\begin{equation}
\label{om-ham}
H_{{\rm om}}=\hbar g(b^\dagger a_1^\dagger a_2+ba_1 a_2^\dagger),
\end{equation}
where we have assumed the resonance condition $\omega_2=\omega_1+\omega_m$. We further assume that the cavity modes are coupled to a single input/output channel. 

A nice feature of this model is that the optomechanical interaction can be configured to turn the mechanical element into a Raman quantum memory by choosing cavity-1 to be driven by a strong coherent pulse.  In the scheme presented here we can exploit this feature to prepare the mechanical system in various states, for example, Fock states or coherent states of varying amplitude. This mechanical coherent state preparation is explained in more detail in Appendix A. 
 
 Once a mechanical state,  say a coherent state, has been loaded into the mechanics, we can then inject single photon states into the optical cavities. This interaction is then described as a controlled beam splitter interaction between the optical modes controlled by the quantum state previously stored in the mechanics. We then repeat this process to do the interferometry so that, at each step, before injecting the single photons, we need to reset the mechanics in vacuum state (by active cooling) and then load it with a coherent state. As reported in the state of the art experiments with PhC optomechanical systems, the mechanical thermalization rate, $\gamma_m \bar{n}_m$, is three orders of magnitude slower than the optical damping rate, $\kappa$~\cite{Chan-cooling1,chan2012optimized}. Therefore, at each trial, we can assume that we detect the photons before the mechanical resonator is damped and we neglect the mechanical damping in this work. 

The total irreversible dynamics of the optomechanical  system is given by the master equation
\begin{equation}
\frac{d\rho}{dt} = -ig[a_1^\dagger a_2  b^\dagger +a_1 a_2^\dagger  b, \rho]+\kappa_1{\cal D}[a_1]\rho+\kappa_2{\cal D}[a_2]\rho,
\end{equation}
where the superoperator ${\cal D}$ is defined by
\begin{equation}
{\cal D}[A]\rho=A\rho A^\dagger-\frac{1}{2}(A^\dagger A\rho+\rho A^\dagger A).
\end{equation}
The respective input and output fields for each cavity are related to the intra-cavity fields by 
\begin{equation}
a_{j,{\rm out}}(t) =\sqrt{\kappa_j}a_j(t)-a_{j,{\rm in}}.
\label{in-out}
\end{equation}

If the mechanics begins in a coherent state $|\beta\rangle$, we can make a canonical transformation (a displacement of the mechanics amplitude) as in section \ref{CPC} to obtain 
\begin{equation}
\frac{d\rho}{dt} = -i\bar{g}[a_1^\dagger a_2  +a_1 a_2^\dagger , \rho]-ig[a_1^\dagger a_2  {\bar b}^\dagger +a_1 a_2^\dagger  {\bar b}, \rho]+\kappa_1{\cal D}[a_1]\rho+\kappa_2{\cal D}[a_2]\rho.
\end{equation}
One might worry if it is valid to treat the dissipative terms for the field as if there was no coherent interaction when $\bar{g}>>1$. The coherent interaction will lead to normal mode splitting of the cavity fields which can indeed alter how they are coupled to the dissipative environment. However, if the local cavity modes are coupled to independent baths (as we assume), with no cross correlations, also flat enough spectral density, and $\kappa_1\approx \kappa_2$, the normal modes are damped at the same rate as the local modes. 

The mechanical system can be prepared in different quantum states by using the mechanical subsystem, as a kind of Raman quantum memory for light. We will assume that, before every preparation step, active cooling is used to prepare the mechanics in the ground state. A strong optical field pulse is directed into the input of one of the cavities to give a good beam splitter interaction between the other cavity and the mechanical element. 

A strong optical coherent field pulse is directed into one of the input optical waveguides, say mode-1, to implement a beam splitter interaction between mode-2 and the mechanical mode. For example, a coherent pulse on the input to cavity-2 can then be transferred to a coherent excitation of the mechanics while a single photon pulse input to cavity-2 will be stored as a Fock state in the mechanics. In this protocol, the mechanical degree of freedom is acting as a quantum memory~\cite{Nunn2008}. We thus have the ability to explore the transition from quantum to classical control described in the first part of this paper. Further details on this preparation stage are discussed in Appendix A.

\section{Mechanically controlled interferometry}
\label{mech-intfr}
Our objective here is to configure the optomechanical system to act as a controlled beam splitter in an interferometer. 
We will assume that the mechanical system has been prepared in a coherent state $|\beta\rangle_b$ (see Appendix A). 
We will then take the input fields $a_{j,{\rm in}}(t)$ to be multi-mode single photon states and perform optical interferometry via a MZ interferometer or HOM interferometer, each using a mechanically controlled beam splitter in place of a conventional beam splitter. These states are injected into one or both of the input modes to each cavity depending the kind of interferometer (MZ or HOM). 
Thus the total initial state is
\begin{align}
 |\psi(0)\rangle &= |0\rangle_1|0\rangle_2|\beta\rangle_b |1_\xi\rangle_{1,{\rm in}} |0\rangle_{2,{\rm in}} \quad ({\rm MZ}),\\
 |\psi(0)\rangle &= |0\rangle_1|0\rangle_2|\beta\rangle_b |1_\xi\rangle_{1,{\rm in}} |1_\eta\rangle_{2,{\rm in}} \quad ({\rm HOM}).
\end{align}
where $|\psi \rangle_{i,{\rm in}}$ is the state of the input field. 

We now need to find the operating conditions so that the optomechanical system can function as a beam splitter port in an interferometer. 

\subsection{Semiclassical limit: open cavities }

As we demonstrated in section \ref{semiclassical}, the semiclassical limit is obtained when the mechanics is prepared in a  coherent state with large coherent amplitude and the coupling constant, $g$, is small, while the effective coupling, $\bar{g}=\beta g$, is constant. We now consider this limit for the case of cavities driven by external single photon sources. We can then compute the visibility of one and two-photon interferometry and how it depends on dephasing corrections that appear in equation (\ref{dephasing}) due to entanglement between the optical and mechanical sub-systems.

Assume that each cavity is driven by single photon pulse states with wavepacket envelope functions $\xi(t),\eta(t)$ for the input to cavity $a_1$ and $a_2$, respectively, and that the coupling between the cavities is given by the first term in the Hamiltonian in equation (\ref{displaced-ham}), the semiclassical approximation. Note that we will assume that the carrier frequency of each single photon pulse is resonant with the respective cavity into which it is injected. We do not explicitly see the carrier frequencies here as we are already working in an interaction picture. We further assume the symmetric case for which $\kappa_1=\kappa_2=\kappa$. In the semiclassical regime, we use the quantum Langevin equations for the optical fields~\cite{QObook,Qnoise}
    \begin{eqnarray}
 \frac{da_1(t)}{dt}=-i\bar g a_2(t)-\frac{\kappa}{2}a_1(t)+\sqrt{\kappa}a_{1,{\rm in}}(t) , \nonumber \\
 \frac{da_2(t)}{dt}=-i\bar g a_1(t)-\frac{\kappa}{2}a_2(t)+\sqrt{\kappa}a_{2, {\rm in}}(t) , 
 \label{Langevin}
   \end{eqnarray}
in which the amplitude functions for single photon state inputs in $a_{1,{\rm in}}(t)$ and $a_{2,{\rm in}}(t)$ are respectively given by $\xi(t)$ and $\eta(t)$. The solution to these linear equations is
\begin{eqnarray}
 a_1(t)&&=\sqrt{\kappa}\Big[A(t)\int_0^t dt^{\prime}\Big(C(t^\prime)a_{1,{\rm in}}(t^\prime)+D(t^\prime)a_{2,{\rm in}}(t^\prime)\Big)\nonumber \\ &&+B(t)\int_0^t dt^{\prime}\Big(D(t^\prime)a_{1,{\rm in}}(t^\prime)+C(t^\prime)a_{2,{\rm in}}(t^\prime)\Big)\Big],\nonumber \\
 a_2(t)&&=\sqrt{\kappa}\Big[B(t)\int_0^t dt^{\prime}\Big(C(t^\prime)a_{1,{\rm in}}(t^\prime)+D(t^\prime)a_{2,{\rm in}}(t^\prime)\Big)\nonumber \\&&+A(t)\int_0^t dt^{\prime}\Big(D(t^\prime)a_{1,{\rm in}}(t^\prime)+C(t^\prime)a_{2,{\rm in}}(t^\prime)\Big)\Big],
 \label{Langevin_sol}
\end{eqnarray}
where $A(t)=e^{-\kappa t/2}\cos (\bar{g}t)$, $B(t)=-ie^{-\kappa t/2}\sin (\bar{g}t)$, $C(t)=e^{\kappa t/2}\cos (\bar{g}t)$ and $D(t)=ie^{\kappa t/2}\sin (\bar{g}t)$. 

We can now derive the effective transmission and reflection coefficients when only one photon is incident on the system. These are defined by 
   \begin{eqnarray}
 R= \int_0^{\infty} \langle a^\dagger_{1, {\rm out}}a_{1,{\rm out}}\rangle_t dt, \nonumber \\
 T=\int_0^{\infty} \langle a^\dagger_{2,{\rm out}}a_{2,{\rm out}}\rangle_t dt. 
  \label{TR}
   \end{eqnarray}
  
\begin{figure}[!htbp]
 \includegraphics[width=14cm]{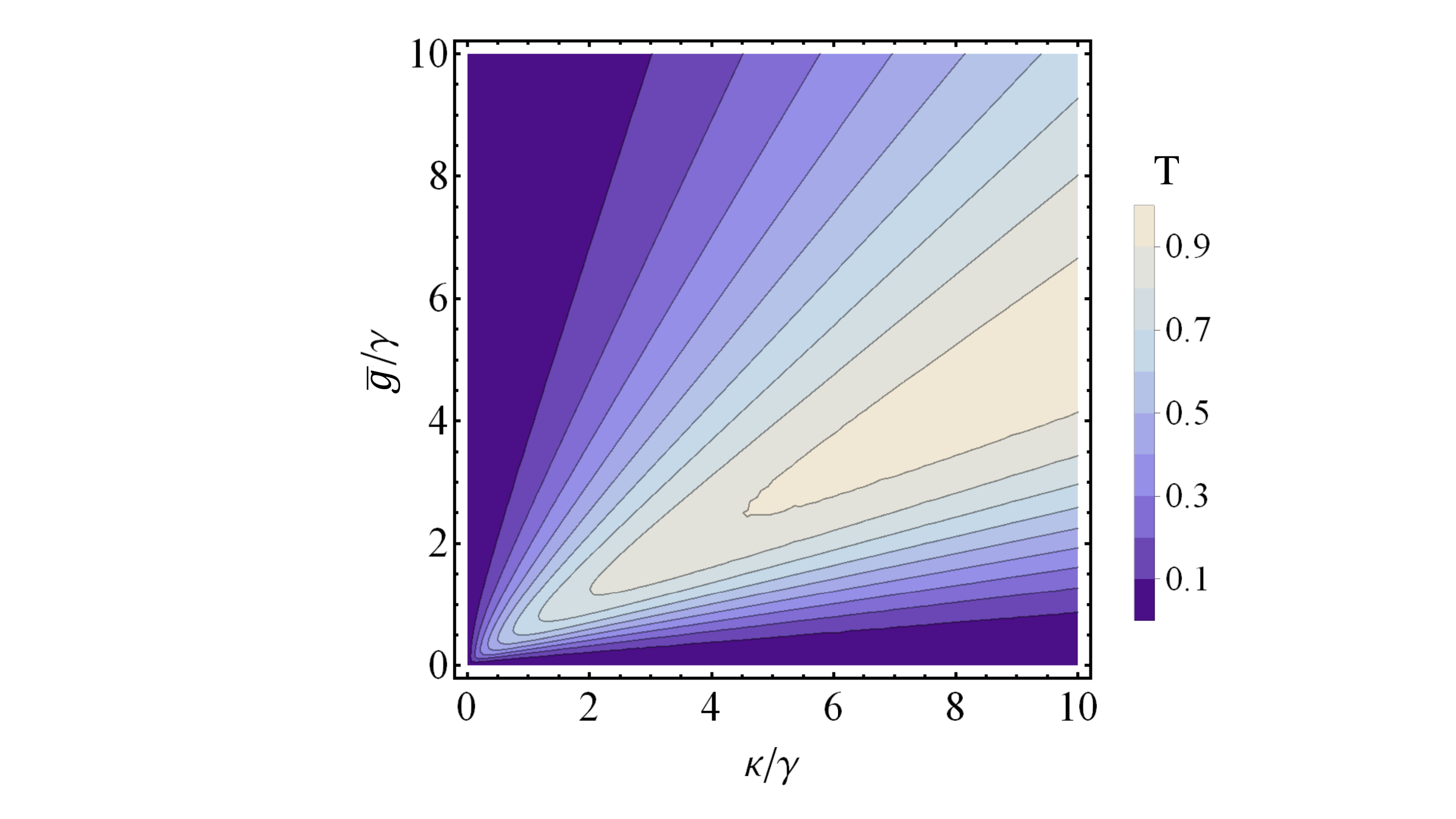}
\caption{The transmission coefficient $T$ versus optical cavity damping rate over the input photon bandwidth ($\kappa/\gamma$) and cavity coupling rate over the input photon bandwidth ($\bar{g}/\gamma$).} 
\label{fig1}
\end{figure}

For the single photon with pulse shape $\xi(t)= \sqrt{\gamma}e^{-\gamma t/2}$, resulting from the decay of a photon from a cavity, these are given by
\begin{eqnarray}
  T=\frac{8 \kappa \bar{g}^2 (\gamma+2\kappa)}{(4\bar{g}^2+\kappa^2)(4\bar{g}^2+(\gamma+\kappa)^2)}, \nonumber \\
  R=1-\frac{8 \kappa \bar{g}^2 (\gamma+2\kappa)}{(4\bar{g}^2+\kappa^2)(4\bar{g}^2+(\gamma+\kappa)^2)},
  \label{TR}
\end{eqnarray}
where $\gamma$ is the single photon bandwidth. Figure \ref{fig1} shows the transmission, $T$, as a function of $\kappa$ and $\bar{g}$ in units of $\gamma$. We can use this figure to configure the optomechanical system as one port in an optical interferometer. Of the two branches in the figure, the lower branch is of experimental relevance  as it enables us to operate with smaller values of  $\bar{g}$. 
\begin{figure}[h!] 
   \centering
   \includegraphics[scale=0.3]{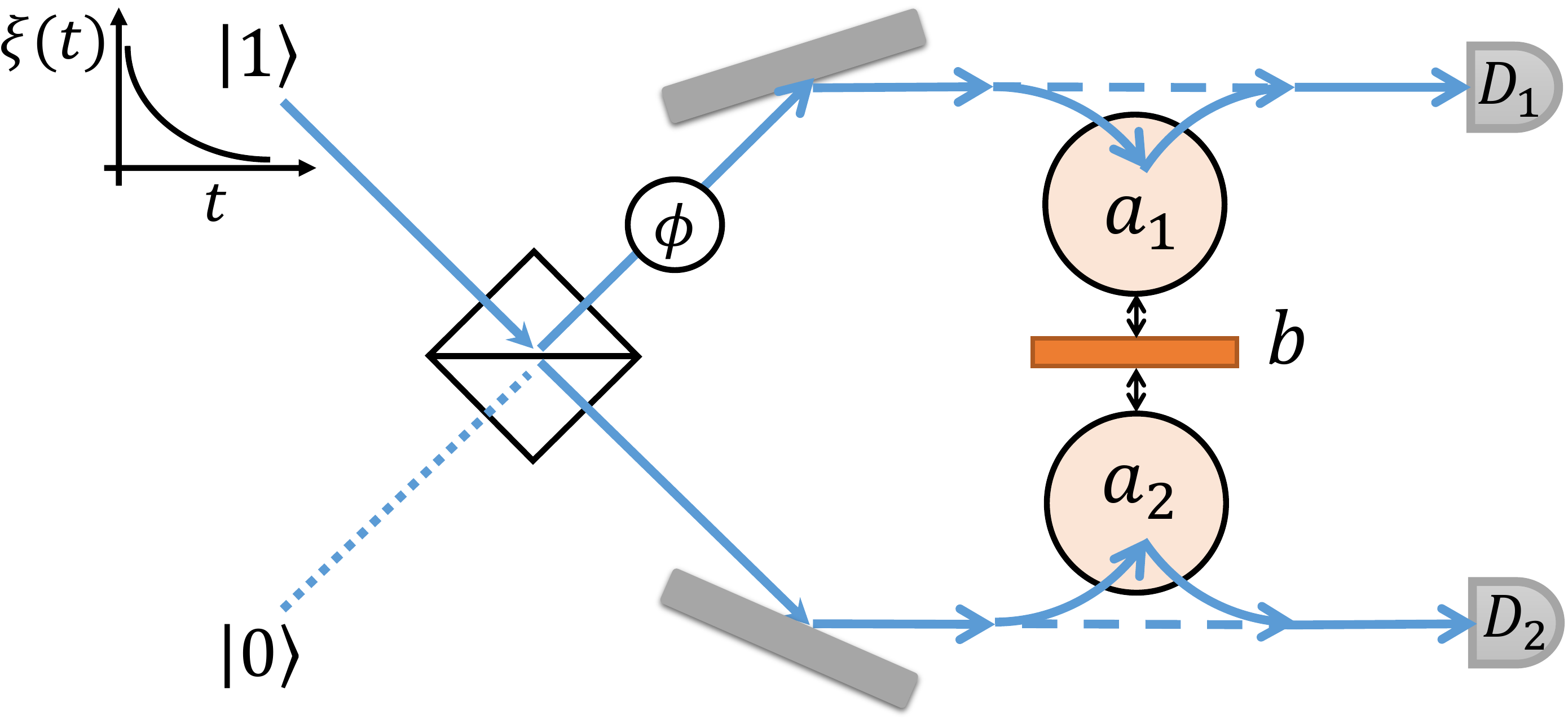} 
   \caption{Scheme for Mach-Zender interferometer in which the first beam splitter is a conventional 50/50 beam splitter and the second beam splitter is replaced by a controlled beam splitter. There is a phase shifter, shown by $\phi$, on the upper arm and we load the interferometer with an exponentially decaying single photon wave packet.}
   \label{fig2}
   \end{figure}

 \subsection{One photon interferometry: Mach-Zender interferometer}
 
 We insert a controlled beam splitter of the type described before by Hamiltonian (\ref{interaction-Hamiltonian}) into the output beam splitter of a MZ interferometer, see figure \ref{fig2}. We inject a single photon with an exponentially decaying shape, $\xi(t)=\sqrt{\gamma} e^{-\frac{1}{2}\gamma t}$, into the interferometer through the port containing cavity-1. Note that the carrier frequency of this pulse needs to be set equal to the resonance frequency of cavity-1.  
 \begin{figure}[h!] 
   \centering
   \includegraphics[scale=0.6]{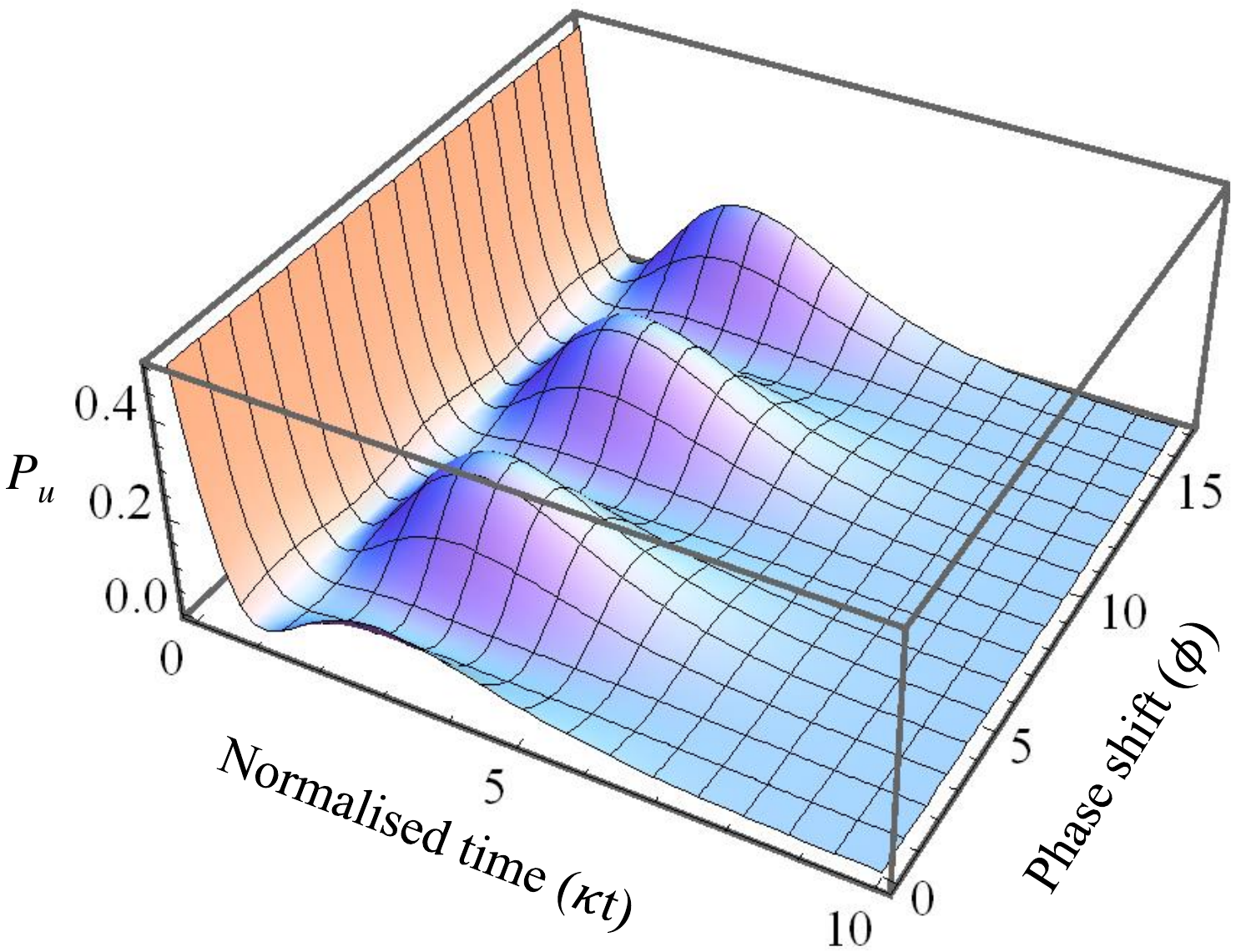} 
   \caption{Probability of detecting the photon at the upper detector, $D_1$, versus phase shift caused in one of the interferometer's arms ($\phi$) and the normalised detection time ($\kappa t$), in the semiclassical regime ($g\rightarrow 0$, $\beta \rightarrow\infty$) for $\kappa=\gamma=1$ and $\bar g=1/3$ which give a transmission of $T=0.4$. The figure shows the interference pattern created at the output of the Mach-Zender interferometer.  }
   \label{mach_3D}
\end{figure}
We further use dimensionless units assuming the cavity damping rate $\kappa=1$ in numerical simulations.
The initial input state to the optomechanical beam splitter after passing the first beam splitter is 
  \begin{equation}
 \vert \psi (0)\rangle=\frac{1}{\sqrt{2}}(e^{i\phi}\vert 1_{a_1},0_{a_2}\rangle+\vert 0_{a_1},1_{a_2}\rangle).
 \label{initial_state}
 \end{equation}
 We use the solutions to the Langevin equations (\ref{Langevin}) given in relations (\ref{Langevin_sol}) to calculate the detection probability in $t$ to $t+dt$ at the upper detector $D_1$ 
 \begin{equation}
 \label{P_u}
 P_u(t:t+dt)=\langle a_{1,{\rm out}}^\dagger(t) a_{1,{\rm out}}(t)\rangle dt.
 \end{equation}
This probability versus detection time and phase shift is plotted in figure \ref{mach_3D}. For $\kappa t\le 1$ the decrease in $P_u$ is due to the transmission of a photon which has not interacted strongly with the mechanics. This is evident because the decay of $P_u$ follows $\xi(t)\propto e^{-\gamma t/2}$. In figure \ref{mach_3D}, it appears that the maximum visibility of the fringes occurs at $\kappa t\approx 3$, but this can be deceiving. For this reason we use $P_u$ to calculate the visibility of the interference pattern at each detection time which is given by
\begin{equation}
v(t)=\frac{P_u^{\rm max}(t)-P_u^{\rm min}(t)}{P_u^{\rm max}(t)+P_u^{\rm min}(t)},
\label{vis}
\end{equation}
where $P^{\rm max}_u$ ($P^{\rm min}_u$) is maximized (minimized) over the phase shift $\phi$.

For the fully quantum mechanical description of the system, we use the unconditional Fock state master equation method \cite{fock_josh,fock_zoller} which for a system having two input modes is
\begin{eqnarray}
\frac{d}{dt}\rho_{m,n;p,q}(t) & = & -i[H,\rho_{m,n;p,q}]+({\cal L}[L_1]+{\cal L}[L_2])\rho_{m,n;p,q}\nonumber \\
&& +\sqrt{m}\xi (t)[\rho_{m-1,n;p,q},L_1^\dagger] +\sqrt{p}\eta (t)[\rho_{m,n;p-1,q},L_2^\dagger] \nonumber \\
&& +\sqrt{n}\xi^* (t)[L_1,\rho_{m,n-1;p,q}]+\sqrt{q}\eta^* (t)[L_2,\rho_{m,n;p,q-1}],
\label{Fock-SME}
\end{eqnarray}
where $H$ is the Hamiltonian given in the equation (\ref{interaction-Hamiltonian}), $L_i=\sqrt{\kappa_i}a_i$ and the superoperator $\cal L$ is defined by
\begin{equation}
{\cal L}[L]\rho=L\rho L^\dagger-\frac{1}{2}(L^\dagger L\rho+\rho L^\dagger L).
\end{equation}

The dynamics is reduced to solving the hierarchy of equations for the operators $\rho_{m,n;p,q}$. These act on the joint Hilbert space of the system and the input fields.   The subscripts $m,n$ refer to the Fock basis for the input to cavity-1 described by the wave packet $\xi(t)$ while $p,q$ refer to the  Fock basis of the input to cavity-2  described by the wave packet $\eta(t)$. As each input has, at most, one photon, the indices are restricted to the values $1,0$. For example, if we had a single photon input at each cavity we would need to solve for $d\rho_{1,1;1,1}$ which couples all the way down to $d\rho_{0,0;0,0}$ in the hierarchy of coupled differential equations. 

After the photon passes through the first beam splitter and the phase shifter, the initial state of the input field incident on the controlled beam splitter is given by equation ({\ref{initial_state}) which is not a pure Fock state. Therefore, the initial sate of the field has the form
\begin{equation}
\rho_{\rm field}(0)=\sum_{m,n,p,q=0}^\infty c_{m,n;p,q}\vert n_{\xi};q_{\eta}\rangle\langle m_{\xi};p_{\eta}\vert,
\label{initialfield}
\end{equation}

The initial conditions are $c_{1,1;0,0}=c_{0,0;1,1}=\frac{1}{2}$, $c_{0,1;1,0}=\frac{1}{2}e^{-i\theta}$, $c_{1,0;0,1}=\frac{1}{2}e^{i\theta}$ and all other coefficients are zero. The initial total state is given by equation (\ref{initialfield}), as~\cite{fock_josh}
\begin{equation}
\rho_{\rm system}(t)=\sum_{m,n,p,q=0}^\infty c_{m,n;p,q}^*\rho_{m,n;p,q}(t).
\label{totalstate}
\end{equation}

Therefore, we solve the hierarchy of differential equations produced by master equation (\ref{Fock-SME}) for $\xi (t)=\eta (t)=\sqrt{\gamma} e^{-\frac{1}{2}\gamma t}$. We need the solutions for $\rho_{1,1;0,0}(t),\rho_{0,0;1,1}(t),\rho_{0,1;1,0}(t)$ and $\rho_{1,0;0,1}(t)$ to calculate the dynamical state of the optomechanical system given by equation (\ref{totalstate}). 
When we use the Fock state master equation approach, a different phase convention to the input-output relation (\ref{in-out}) is used (see section five of \cite{Qnoise}) such that we have
\begin{equation}
a_{j,{\rm out}}(t) =\sqrt{\kappa_j}a_j(t)+a_{j,{\rm in}}(t).
\label{in-out2}
\end{equation}

\begin{figure}[h!] 
   \centering
   \includegraphics[scale=0.4]{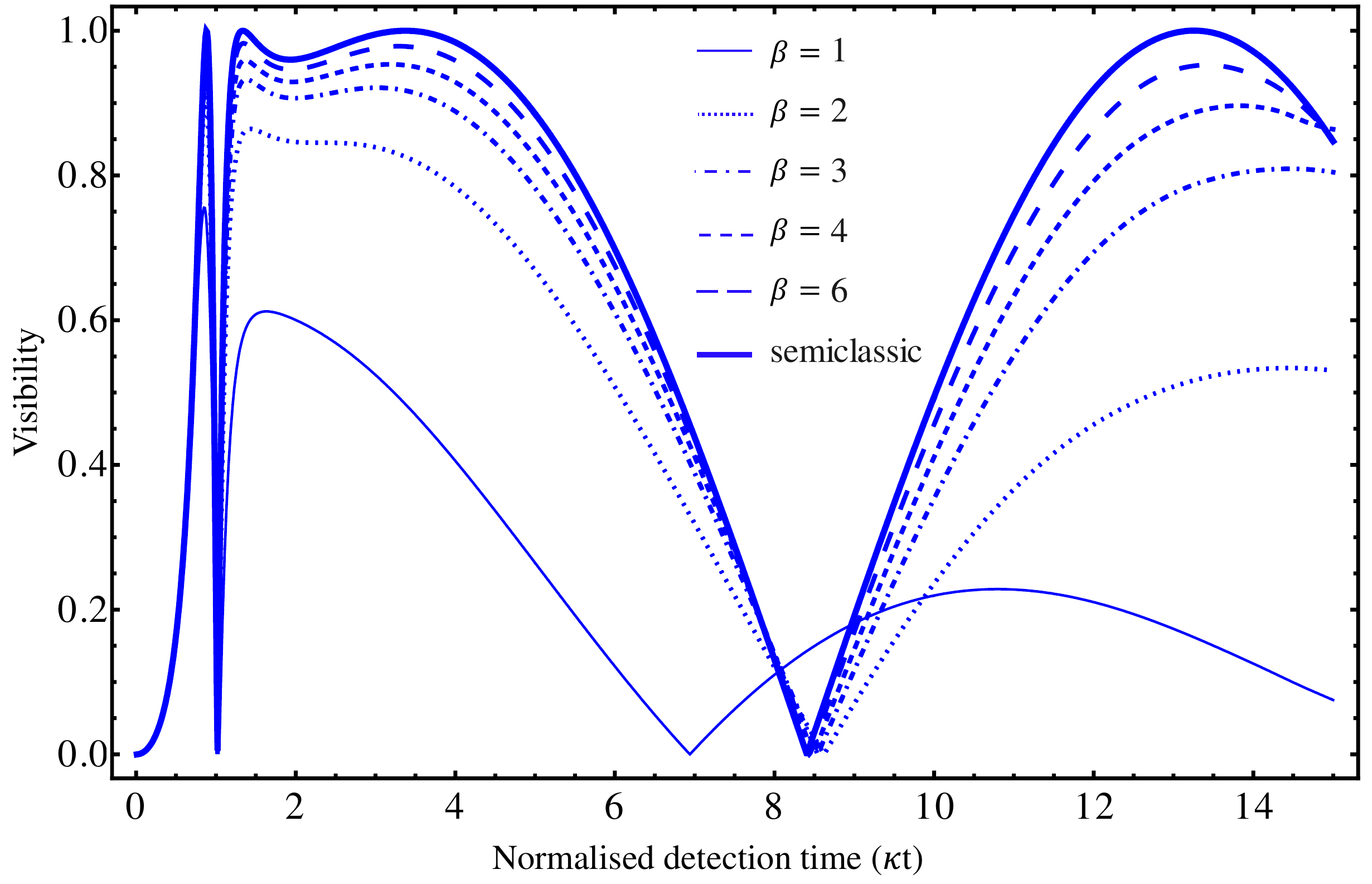} 
   \caption{Interference visibility of a MZ interferometer versus detection time for different values of the mechanical coherent state amplitude for $\kappa=\gamma=1$, $\bar g=1/3$. For large enough mechanical coherent state amplitude, $\beta>6$, visibility transits towards that one obtained in the semiclassical regime. }
   \label{mach_vis}
\end{figure}
  \begin{figure}[h!] 
   \centering
   \includegraphics[scale=0.4]{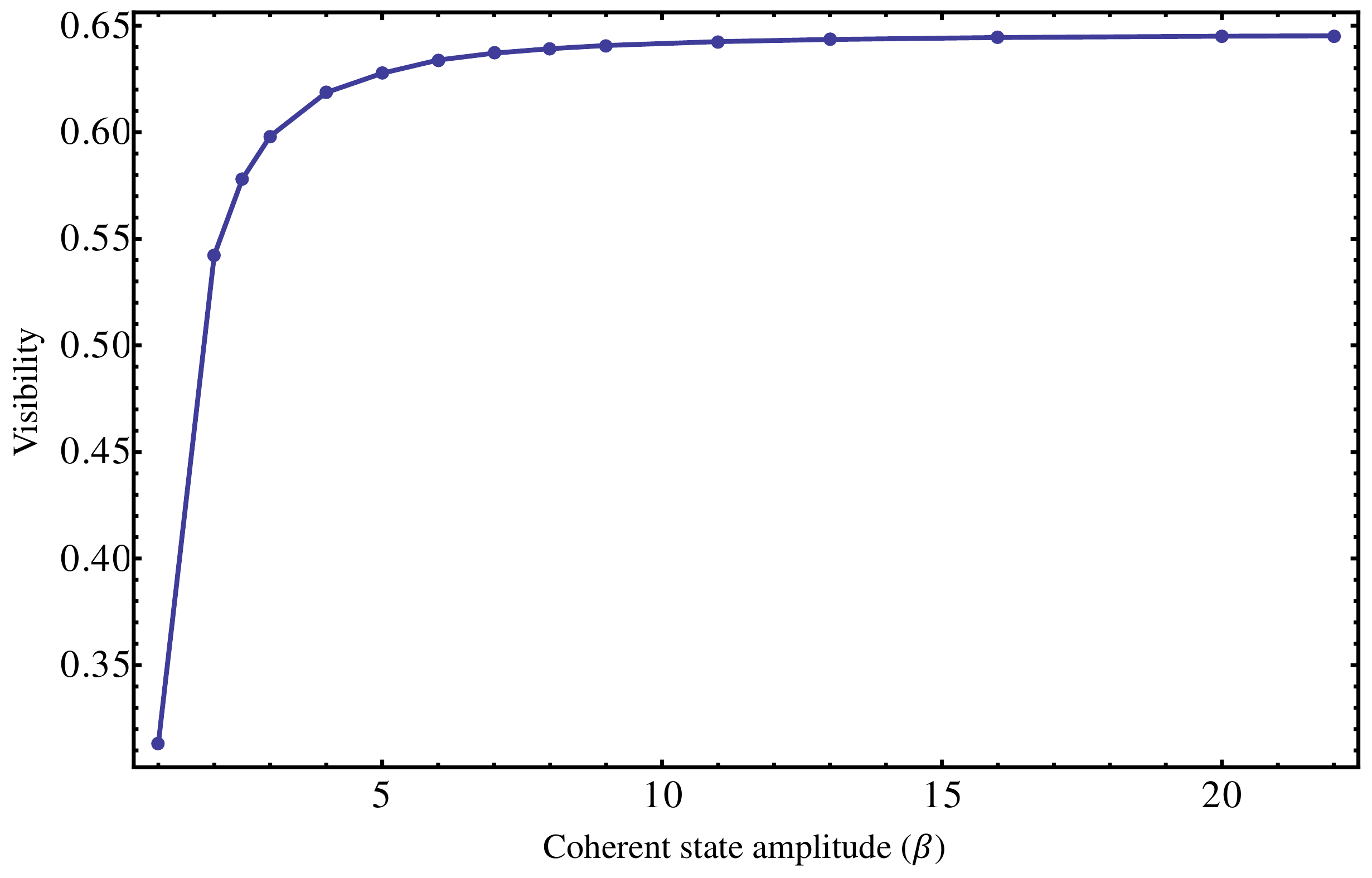} 
   \caption{Interference pattern Visibility versus mechanical coherent state amplitude for $\kappa=\gamma=1$, $\bar g=1/3$. With these parameters, the visibility obtained for the effective beam splitter in the semiclassical regime is 0.646.}
   \label{fig5}
\end{figure}

To calculate the detection probability at the top detector, $D_1$, which is defined in equation (\ref{P_u}), one also needs the action of the operators $a_{1,{\rm in}}(t)$ and $a_{2,{\rm in}}(t)$ on two mode Fock states
\begin{eqnarray}
&&a_{1,{\rm in}}(t)\vert n_{\xi};q_\eta\rangle=\xi(t)\sqrt{n}\vert n-1_{\xi};q_\eta\rangle,\nonumber\\
&&a_{2,{\rm in}}(t)\vert n_{\xi};q_\eta\rangle=\eta(t)\sqrt{q}\vert n_{\xi};q-1_\eta\rangle.
\end{eqnarray}
We then use equation (\ref{vis}) to calculate the visibility of the interference pattern for different values of the coherent state amplitude shown in figure \ref{mach_vis}. This figure demonstrates that as $\beta$ increases, the corrections due to the first and higher order terms in $g$, which was discussed in section {\ref{CPC}, become negligible and when $\beta$ is large enough, we recover a semiclassical interaction without any entanglement between the optical and mechanical degrees of freedom.

In practice, photo detector operates with a finite integration time. Therefore, it is more convenient to integrate over time to calculate the detection probability at $D_1$ as 
 \begin{equation}
 P_u=\int_0^{\infty}\langle a_{1,{\rm out}}^\dagger(t) a_{1,{\rm out}}(t)\rangle dt,
 \end{equation}
which gives a detection probability and an interference visibility independent of the detection time as one would expect in an experiment. This visibility is plotted in figure \ref{fig5} versus coherent state amplitude prepared in the mechanics. This figure shows that by enhancing $\beta$, the visibility saturates to the value obtained in the semiclassical limit. The interference visibility of a MZ interferometer can be used as a sign to show the transition from quantum control to classical control that is obtained for a certain value of $\beta$.

\subsection{Hong-Ou-Mandel interferometer}
The scheme for our system working as a beam splitter inside a HOM interferometer is shown in figure \ref{fig6}. We send one photon to each of the modes $a_1$ and $a_2$. The single photons are specified by the same amplitude function except one is time shifted with respect to the other
\begin{align}
\xi(t)&=\sqrt{\gamma} e^{-\frac{1}{2}\gamma t},\nonumber \\
\eta(t)&=\sqrt{\gamma} e^{-\frac{1}{2}\gamma (t-\tau)},
\end{align}
and, as before, we assume that they have carrier frequencies resort with their respective cavities. 
We can then use differential equations (\ref{Langevin}) together with the input-output relation (\ref{in-out}) to analytically calculate the probability of the joint photon counting at $D_1$ and $D_2$ in the semiclassical regime~\cite{BasiriOpx} which is defined as 
\begin{eqnarray}
G^{(2)}(\tau)=\frac{\int_0^\infty \int_0^\infty \langle a_{1,{\rm out}}^\dagger (t)a_{2,{\rm out}}^\dagger (t')a_{2,{\rm out}}(t')a_{1,{\rm out}}(t)\rangle dtdt'}{\int_0^\infty\langle a_{1,{\rm out}}^\dagger a_{1,{\rm out}}(t)\rangle dt\int_0^\infty\langle a_{2,{\rm {\rm out}}}^\dagger a_{2,{\rm out}}(t)\rangle dt}.
\label{G2}
\end{eqnarray}
\begin{figure}[h!] 
   \centering
   \includegraphics[scale=0.3]{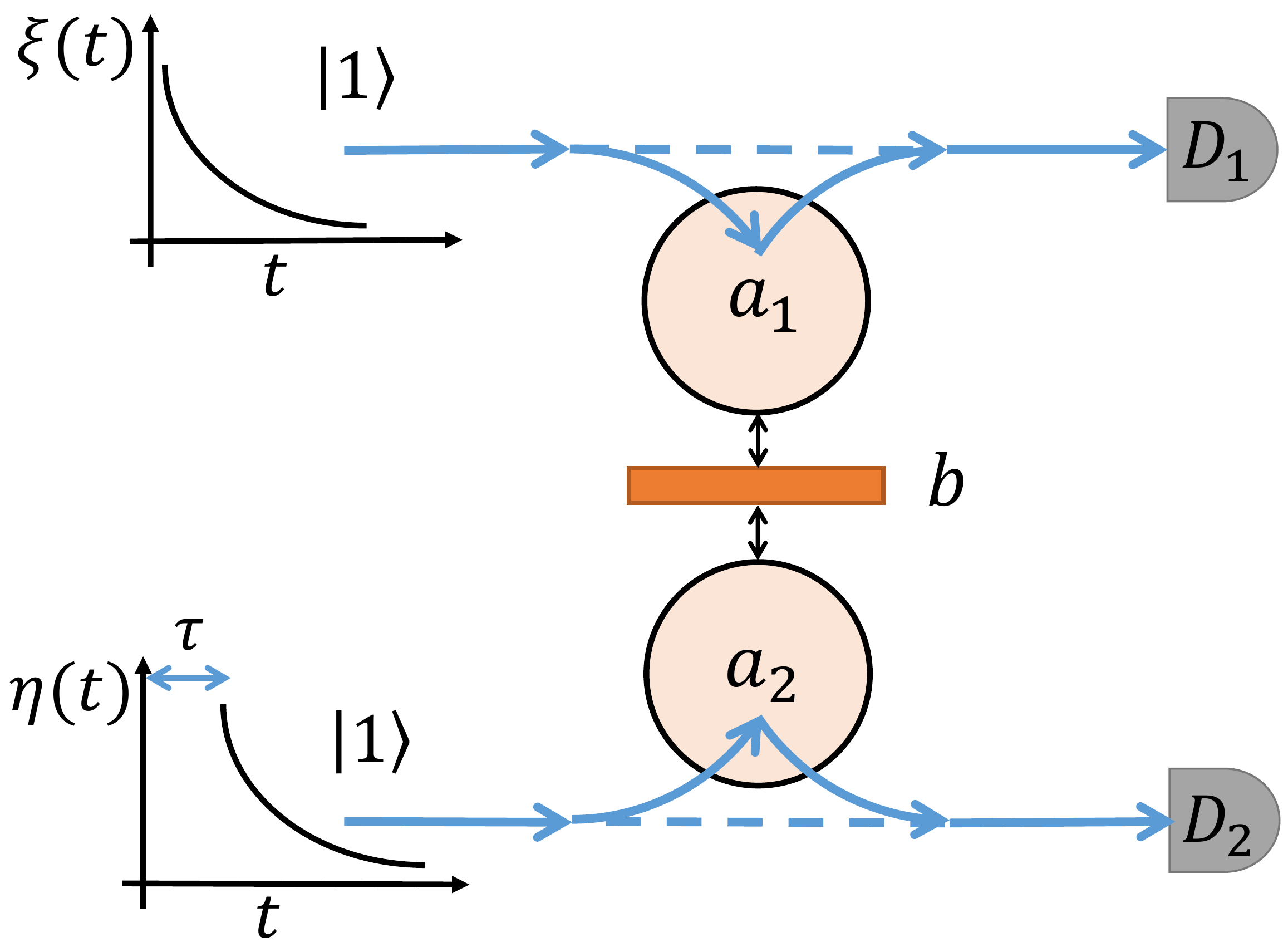} 
   \caption{Scheme for Hong-Ou-Mandel interferometer using a mechanically controlled beam splitter. Input single photons have exponentially decaying pulse shapes entering the optical cavities with a time shift $\tau$.   }
   \label{fig6}
\end{figure}
In the quantum regime, one way to calculate this two-time correlation function is to use the quantum regression theorem, for which one needs to solve the unconditional master equation. However, for this two-mode, two-input photon case, this becomes complicated. Another way is to numerically simulate a HOM experiment using the stochastic theory of quantum jumps. We choose the latter approach in this work, the details for which will be given later in this section. Before that, we start with a simpler calculation to give some physical insight into testing the quantum to classical control by employing a HOM interferometer.

We consider the physically idealistic case of coincidence detection at a specific detection time $t$ by calculating  the coincidence detection rate $C(t,\tau)=\kappa_1\kappa_2 \langle a_1(t)^\dagger a_1(t)a_2(t)^\dagger a_2(t)\rangle$. This expectation value can be computed using the Fock state master equation (\ref{Fock-SME}) as
\begin{equation}
C(t,\tau)=\kappa_1\kappa_2 {\rm Tr}[a_1^\dagger a_1a_2^\dagger a_2\rho_{1,1;1,1}(t)].
\end{equation}
This coincidence rate is plotted in figure \ref{fig7}(a) versus detection time and the time shift between the input photons for the semiclassical regime. However, we get qualitatively the same plot for the quantum regime with a HOM dip forming around the $\tau=0$ point. The HOM dip can be clearly seen in this figure at fixed detection times. We choose $\kappa t=4.7$ for which the HOM visibility in the semiclassical regime is 1 and then plot the coincidence rate versus the time shift between the input photons for different values of coherent state amplitude, $\beta$, changing from a quantum regime, $\beta=1$, to very strong amplitudes, as can be seen in figure \ref{fig7}(b). 
\begin{figure} 
   \centering
\includegraphics[scale=0.6]{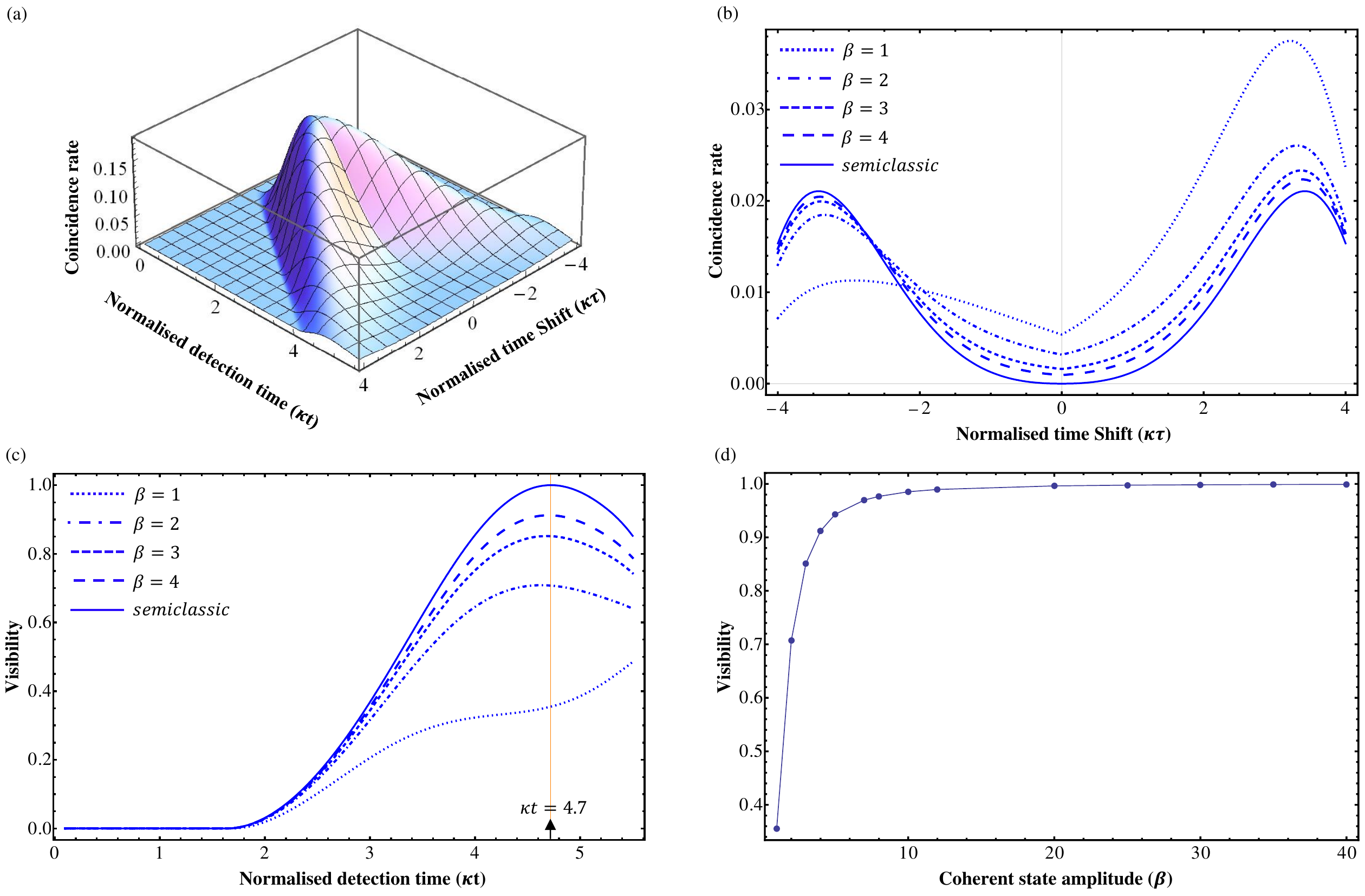} 
   \caption{In these figures $\kappa=\gamma=1$, $\bar g=1/3$. (a) Coincidence rate versus detection time ($\kappa t$) and the time shift between the entering photons for the mechanical state in the semiclassical regime. As the figure shows, the HOM dip changes for different interaction times. (b) HOM dip for different values of initial coherent state amplitudes. As $\beta$ becomes close to the semiclassical limit the asymmetry in the HOM dip disappears. (c) Visibility versus detection time for $\beta$ ranging from small values in fully quantum regime to larger values in the semiclassical regime. (d) HOM interference visibility for different values of initial coherent state amplitudes ($\beta$) at $\kappa t=4.7$. }
   \label{fig7}
\end{figure}
One feature we observe in this figure is the asymmetry in the HOM dip in the quantum regime, which arises from the asymmetry in interaction Hamiltonian given in equation (\ref{interaction-Hamiltonian}). By increasing $\beta$, this asymmetry is gradually removed since as we increase $\beta$, there is a gradual transition to the semiclassical
regime with a symmetric interaction Hamiltonian between the optical modes. This figure also suggests that the change in the HOM dip can be used as a measure of the transition from the quantum control regime to the classical control regime.

We calculate the visibility of the HOM interference pattern as 
\begin{equation}
v(t)=\frac{C^{\rm max}(t,\tau_{\rm negative})-C(t,0)}{C^{\rm max}(t,\tau_{\rm negative})+C(t,0)},
\end{equation}
which gives the worst case visibility for each curve. Figure \ref{fig7}(c) shows HOM visibility versus the detection time. For the chosen parameter regimes of $\gamma$ and $\bar g$ in this figure, maximum visibility occurs at $\kappa t=4.7$, the detection time that we choose to plot visibility versus coherent  state amplitude prepared in the mechanics in figure \ref{fig7}(d). This figure also shows that HOM visibility can be an indicator to test the transition form the quantum control to the classical control in which the visibility saturates to a maximum value. 

In the next step, we perform the calculation using the more realistic definition of the joint detection probability given in equation (\ref{G2}). As discussed earlier, we perform a Monte-Carlo simulation using the stochastic version of the Fock state master equation to simulate the HOM interference which shows the ratio of coincident photo-detections over the total number of measurements versus the time shift, ($\tau$). We need to model the conditional evolution of the system. Following the same procedure as introduced for unconditional Fock state master equation~\cite{fock_josh}, we derive the conditional master equation (for more details see ~\cite{BaraCom14}) describing the dynamics of the system given vacuum detection in both modes up to time $t$ 
\begin{eqnarray}
\frac{d}{dt}\tilde{\rho}_{m,n;p,q}^{(0_1,0_2)}(t) & = & -i[H,\rho_{m,n;p,q}]-\frac{1}{2}\lbrace L_1^\dagger L_1,\rho_{m,n;p,q}\rbrace-\frac{1}{2}\lbrace L_2^\dagger L_2,\rho_{m,n;p,q}\rbrace\nonumber \\
&& -\sqrt{m}\xi (t)L_1^\dagger\rho_{m-1,n;p,q} -\sqrt{p}\eta (t)L_2^\dagger\rho_{m,n;p-1,q} \nonumber \\
&& -\sqrt{n}\xi^* (t)\rho_{m,n-1;p,q}L_1-\sqrt{q}\eta^* (t)\rho_{m,n;p,q-1}L_2\nonumber\\ &&-\sqrt{mn}\vert\xi(t)\vert^2\rho_{m-1,n-1;p,q}-\sqrt{pq}\vert\eta(t)\vert^2
\rho_{m,n;p-1,q-1},
\label{cond_SME}
\end{eqnarray}
where $\tilde{\rho}_{m,n;p,q}^{(0_1,0_2)}(t)$ is the conditional un-normalised state of the system in which $n_i$ in the superscript $(n_1,n_2)$ is the number of counts at detector $D_i$. The top level generalized density operator $\tilde{\rho}_{1,1;1,1}^{(0_1,0_2)}(t)$ is the physical state of the system and is used to calculate the normalization factor: ${\rm Tr}[\tilde{\rho}_{1,1;1,1}^{(0_1,0_2)}(t+dt)]$, which is in fact the probability for a vacuum detection occurring in
the time interval $(t,t+dt]$ 
\begin{eqnarray}
P^{(0_1,0_2)}(t)& = &1-dt\Big(\vert\xi(t)\vert^2 {\rm Tr}[\rho_{0,0;1,1}(t)]-\vert\eta(t)\vert^2 {\rm Tr}[\rho_{1,1;0,0}(t)] \nonumber \\ &&-{\rm Tr}[L_1^\dagger L_1\rho_{1,1;1,1}]-{\rm Tr}[L_2^\dagger L_2\rho_{1,1;1,1}]\nonumber \\ &&-\xi(t){\rm Tr}[L_1^\dagger\rho_{0,1;1,1}]-\xi^*(t){\rm Tr}[L_1\rho_{1,0;1,1}]\nonumber \\ &&-\eta(t){\rm Tr}[L_2^\dagger\rho_{1,1;0,1}]-\eta^*(t){\rm Tr}[L_2\rho_{1,1;1,0}]\Big).
\label{P0}
\end{eqnarray}

The conditional state of the system given that one photon is detected at $D_1$ in the time interval $(t,t+dt]$, should be updated as
\begin{eqnarray}
\tilde{\rho}_{m,n;p,q}^{(1_1,0_2)}(t+dt)& = &dt\Big(L_1\rho_{m,n;p,q}(t)L_1^\dagger+\sqrt{mn}\vert\xi(t)\vert^2\rho_{m-1,n-1;p,q}\nonumber\\ &&+\sqrt{m}\xi(t)\rho_{m-1,n;p,q}L_1^\dagger+\sqrt{n}\xi(t)^*L_1\rho_{m,n-1;p,q}\Big),
\label{J1}
\end{eqnarray}
and for a count occurring at $D_2$ in $t$ to $t+dt$ we have
\begin{eqnarray}
\tilde{\rho}_{m,n;p,q}^{(0_1,1_2)}(t+dt)& = &dt\Big(L_2\rho_{m,n;p,q}(t)L_2^\dagger+\sqrt{pq}\vert\eta(t)\vert^2\rho_{m,n;p-1,q-1}\nonumber\\ &&+\sqrt{p}\eta(t)\rho_{m,n;p-1,q}L_2^\dagger+\sqrt{q}\eta(t)^*L_2\rho_{m,n;p,q-1}\Big),
\label{J2}
\end{eqnarray}
The associated normalization with the states given in equations (\ref{J1}) and (\ref{J2}) gives the probability for a photo-detection occurring in the time interval $(t,t+dt]$ at detectors $D_1$ and $D_2$, respectively
\begin{eqnarray}
P^{(1_1,0_2)}(t)& = &dt\Big({\rm Tr}[L_1^\dagger L_1\rho_{1,1;1,1}(t)]+\vert\xi(t)\vert^2\rho_{0,0;1,1} \nonumber \\&&+\xi(t){\rm Tr}[L_1^\dagger\rho_{0,1;1,1}(t)] +\xi^*(t){\rm Tr}[L_1\rho_{1,0;1,1}(t)]\Big),
\label{P1}
\end{eqnarray}
\begin{eqnarray}
P^{(0_1,1_2)}(t)& = &dt\Big({\rm Tr}[L_2^\dagger L_2\rho_{1,1;1,1}(t)]+\vert\eta(t)\vert^2\rho_{1,1;0,0} \nonumber \\&&+\eta(t){\rm Tr}[L_2^\dagger\rho_{1,1;0,1}(t)] +\eta^*(t){\rm Tr}[L_2\rho_{1,1;1,0}(t)]\Big).
\label{P1}
\end{eqnarray}

We perform the two-jump Monte-Carlo simulation in four steps as follows: (1) start with the optomechanical initial state $|\psi(0)\rangle = |0\rangle_1|0\rangle_2|\beta\rangle_b$ and inject two single photons, with a time shift $\tau$, to the cavity inputs. (2) Generate a random number, $rand$, in the range 0 to 1. If $P^{(0_1,0_2)}(t:t+dt)>rand$, no jump occurs and the normalised state of the system at the end of the interval should be updated as 
\begin{equation}
\rho_{m,n;p,q}(t+dt)=\frac{\tilde{\rho}_{m,n;p,q}^{(0_1,0_2)}(t+dt)}{P^{(0_1,0_2)}(t:t+dt)}.
\end{equation}
If $P^{(0_1,0_2)}(t:t+dt)<rand$, a jump occurs and we choose a second random number $rand_J$ to decide if the jump occurs in mode one or in mode two.
\begin{equation}
{\rm If} \frac{P^{(1_1,0_2)}(t:t+dt)}{1-P^{(0_1,0_2)}(t:t+dt)}>rand_J \longrightarrow \rho_{m,n;p,q}(t+dt)=\frac{\tilde{\rho}_{m,n;p,q}^{(1_1,0_2)}(t+dt)}{P^{(1_1,0_2)}(t:t+dt)}.
\end{equation}
\begin{equation}
{\rm If} \frac{P^{(1_1,0_2)}(t:t+dt)}{1-P^{(0_1,0_2)}(t:t+dt)}<rand_J \longrightarrow \rho_{m,n;p,q}(t+dt)=\frac{\tilde{\rho}_{m,n;p,q}^{(0_1,1_2)}(t+dt)}{P^{(0_1,1_2)}(t:t+dt)}.
\end{equation}
(3) We repeat step 2 until we have detected both photons. (4) We repeat steps 1 to 3 for a large number of trajectories.

\begin{figure}[h!] 
   \centering
   \includegraphics[scale=0.4]{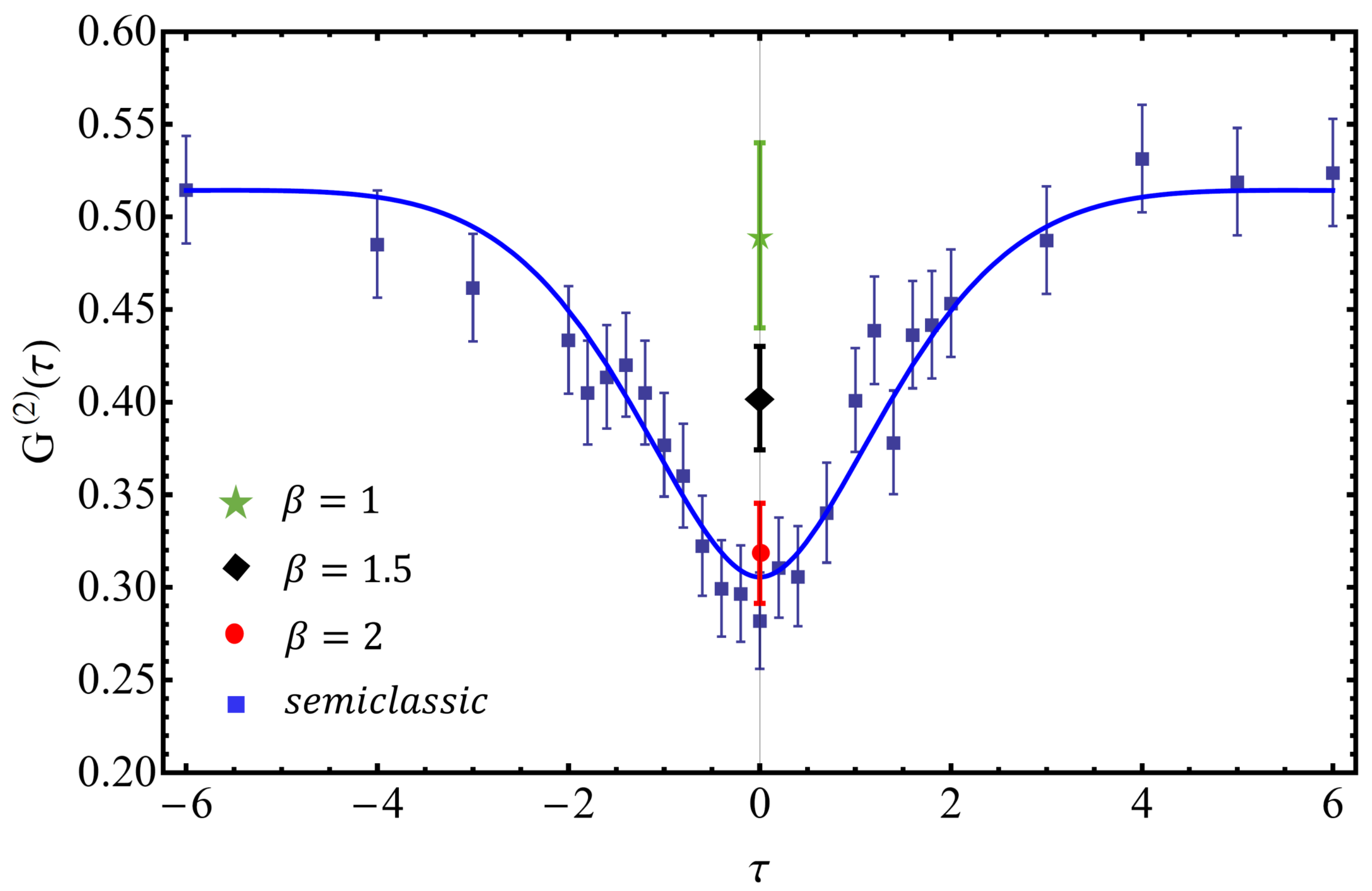} 
   \caption{Joint detection probability versus the time shift between the entering photons to the interferometer for $\kappa=\gamma=1$, $\bar g=1/3$. The solid line shows the analytical results for the semiclassical regime. Data points with error bars are the results achieved by solving the conditional stochastic Fock state master equation in a Monte-Carlo simulation. Each data point is the result of simulations for 1200 trajectories except for $\beta=1$, in which 400 trajectories were performed. In the semiclassical limit we observe a very good agreement between the analytical calculations using Langevin equations and the numeric done using Fock state master equation. The error bars are two standard deviation of a Bernoulli distribution. }
   \label{fig8}
\end{figure}
Figure \ref{fig8} shows the probability of having one count at each of detectors $D_1$ and $D_2$. The solid line shows the analytical results for the semiclassical regime as a solution to equation (\ref{G2})~\cite{BasiriOpx}. Blue squares show the numerical results for the semiclassical case obtained by using Monte-Carlo simulation. For each $\tau$, we performed 1200 trajectories. For other values of $\beta$ in the quantum regime, we only compute the joint detection probability at $\tau=0$ since we are limited by our computation resources. However, this plot clearly shows the trend we expect to see; a decrease in the HOM dip with increasing mechanical coherent state amplitude, as we observed in the previous figures. 

In the semiclassical limit, there is an offset from zero in the HOM dip for totally indistinguishable photons. Considering the fact that with the chosen parameters the effective beam splitter is performing at $T=0.4$, we also give a full description of the behavior of the effective beam splitter in the semiclassical regime implemented in both MZ and HOM interferometers in Appendix B. The analysis fully include all the phenomena involved in the visibility reduction in this effective beam splitter which are not involved in a conventional beam splitter interaction.

\section{Discussion and conclusion}
\label{discussion}
In this paper, we have specified an optomechanical scheme in which the quantum state of a mechanical resonator can be used as a quantum controller for single photon excitations in each of two waveguide modes. The mechanical resonator controls the exchange of photons between two coupled cavities evanescently coupled to optical waveguides. The Hamiltonian of our model realizes a three-wave mixing process via a cubic optomechanical nonlinear interaction. By controlling the quantum state of the mechanical resonator we can realise a quantum controlled beam splitter for the two input optical modes. The model we implement makes extensive use of a recently developed formalism for dealing with non stationary input Fock states to optical cavities and serves a non-trivial application of this tool. 

We have demonstrated that, when the mechanics is prepared in a deep quantum state, the controlled beam splitter instantiates a kind of optical mechanical Stern-Gerlach interferometer for an optical qubit (one-photon case) and qutrit (two-photon case). We have shown how increasing the degree of coherent excitation in the mechanical resonator leads to a decrease in the entanglement between the optical and mechanical degrees of freedom and further show how this may be monitored using HOM interferometry of the input optical photons. In figure \ref{mach_vis}, we see that the {\em minimum visiblity} for normalised detection times  $\kappa t>1$, is sensitive to the amplitude of the mechanics coherent state. Specifically the visibility varies rapidly when the mechanics is acting as a quantum controller i.e. $|\beta|<2$. When $|\beta|\ge 2$ the mechanics is essentially classical. Similar conclusions can be drawn from figure \ref{fig8}. We see $G^{(2)}(0)$ varies rapidly for $|\beta|<2$. Both of these effects are therefore signatures that the mechanics is behaving as a genuine quantum controller.  This decrease in entanglement is explained in terms of the Stern-Gerlach analogy by the gradual loss of which-path information stored in the mechanical resonator. Further, these effects could be used as a weak force sensor if the weak force drives the coherent excitation of the mechanical resonator. Given the ability to prepare the mechanical degree of freedom in coherent states of varying amplitude, this model demonstrates the emergence of classical control from an underlying quantum mechanical model.

\section*{Appendix A: Preparation of the mechanical system}

The mechanical system can be prepared in a coherent state in two ways. In both methods, first a laser cooling scheme prepares the mechanical oscillator in the ground state. First method is to apply a classical resonant force to the mechanics to drive it to a steady state which is a coherent state. Second approach is based on the mechanical degree of freedom operating in a quantum memory mode~\cite{Nunn2008}. In this approach, we drive one of the optical modes by a strong continuous coherent field to implement a beam splitter interaction between the mechanics and the other optical mode. Then we send a strong coherent pulse to the second optical mode. This coherent field can then be transferred to a coherent excitation of the mechanical mode as a result of the beam splitter interaction between the mechanics and optical mode-2. In both approaches, we have the potential to prepare the mechanics in coherent sates with different amplitudes, $\beta$. Below we study the latter method in more details.

Suppose we wish to load a coherent state into the memory. In that case the input to both optical cavities are coherent time dependent pulses. The input pulse to cavity-1 will be taken to be a very strong coherent pulse and we will refer to this as the read/write (R/W) pulse.  The input R/W pulse is assumed to be in a coherent state with complex amplitude ${\cal E}(t)$ which is an external field to the cavity so the pulse intensity, $|{\cal E}(t)|^2$, must have units of flux ($s^{-1}$).  We now make a canonical transformation
\begin{equation}
a_1=\bar{a}_1+\alpha(t),
\label{displacement}
\end{equation}
where $\alpha(t)$ is the time dependent complex field amplitude of the control pulse inside the cavity. The interaction Hamiltonian is then written as
\begin{equation}
\label{H_mech-preparation}
H=\hbar g (a_2b^\dagger\alpha(t)^*+a_2^\dagger b\alpha(t))+\hbar g (\bar{a}_1^\dagger a_2b^\dagger +\bar{a}_1 a_2^\dagger b).
\end{equation} 
In order to operate as a quantum memory we would like R/W cavity field ($a_1$) to respond quickly to the input pulse, ${\cal E}(t)$ so that $\alpha(t)$ is slaved to ${\cal E}(t)$  (the adiabatic approximation)
\begin{equation}
\alpha(t)=\frac{2{\cal E}(t)}{\sqrt{\kappa_1}}.
\label{cavity-amp}
\end{equation}
In order to swap the state of the cavity mode-2 to the mechanics, the strong control pulse should be always on over the time required to write to the memory. In this case, $\alpha(t)$ is very large over the interaction time between the signal and the memory, so with a good approximation we can ignore the second term in equation (\ref{H_mech-preparation}) compared to the first term.

The output fluctuation field (i.e. the output field minus the coherent component) is thus given by
\begin{equation}
a_{1,{\rm out}} =\frac{-2i g}{\sqrt{\kappa}_1} a_2b^\dagger+ a_{1,{\rm in}}.
\end{equation}
If we  assume that $g/\kappa_1 <<1$,  the output R/W field is virtually the same as the input R/W field, i.e. coherent, and the entanglement with the other degrees of freedom can be neglected.  It is possible to account for the residual entanglement between the R/W modes in the adiabatic approximation by  the master equation, in the interaction picture,  
\begin{equation}
\label{adiabatic-me}
\frac{d\rho}{dt} =-ig [a_2b^\dagger\alpha(t)^*+a_2^\dagger b\alpha(t),\rho]+\Gamma{\cal D}[a_2b^\dagger]\rho+\kappa_2{\cal D}[a_2]\rho,
\end{equation}
where 
\begin{equation}
\Gamma =\frac{4g^2}{\kappa_1}.
\end{equation}
The second last term describes a correlated quantum jump via the jump operator $a_2b^\dagger$ wherein the memory is accidentally excited and one photon is absorbed from the cavity mode-2.  In order to use this system as a quantum memory we require that over the time $T$ of the RW pulse 
\begin{equation}
\label{memory-condition1}
\Gamma T << g A,
\end{equation}
where $A= \int_0^T\alpha(t)dt$ is the pulse area, so that we can neglect the residual entanglement described by the second term in equation (\ref{adiabatic-me}) over the times required to write or read to the memory.

We now assume that cavity mode-2 is continuously driven by a coherent driving field, with amplitude $\epsilon$, resonant with the cavity mode. The time dependent interaction Hamiltonian for reading and writing to the memory is 
\begin{equation}
H_m=\hbar g (a_2b^\dagger\alpha^*(t)+a_2^\dagger b\alpha(t))+(\epsilon a_2^\dagger+ \epsilon^*a_2).
\end{equation}
The corresponding quantum stochastic differential equations for the memory are
\begin{eqnarray}
\frac{db}{dt} & = & -ig a_2 \alpha^*(t),\\
\frac{da_2}{dt} & = & -ig\alpha(t) b -\frac{\kappa_2}{2} a_2-i\epsilon +\sqrt{\kappa_2} a_{i,2}.
\end{eqnarray}
Prior to the R/W pulse switching on, the cavity will have reached a steady state which is in fact a coherent state $|\alpha_0\rangle$ with coherent amplitude 
\begin{equation}
\alpha_0=\frac{-2 i\epsilon}{\kappa_2}.
\end{equation}
We define a change of variable according to 
\begin{equation}
\theta(t) =\frac{1}{A}\int_{-\infty}^t \alpha(t')dt'.
\end{equation}
Thus $\theta(t)$ is a sigmoidal function between $0$ and $1$ and centered on the R/W pulse.

 We now assume that the temporal width of the R/W pulse, $T$,  is sufficiently short that $\kappa_2T<<1$. This means that over the time that 
the R/W pulse is significantly different from zero we can neglect the decay of the cavity. In that case, we can approximate the dynamics over the time of the pulse by
\begin{eqnarray}
\frac{db}{d\theta} & = &  -i\tilde{g}a_2,\\
\frac{da_2}{d\theta} & = & -i\tilde{g} b,
\end{eqnarray}
where the dimensionless coupling constant is given by $\tilde{g}=  gA$ and with initial condition set as $|\alpha_0\rangle_2\otimes|0\rangle_b$. The solution to these equations is given by a unitary transformation with generator
$G=\tilde{g}(a^\dagger  b+ab^\dagger)$. If we choose $\tilde{g}\theta=\pi/2$ we find that the initial state thus evolves to 
\begin{equation}
e^{-i\pi G/2}|\alpha_0\rangle_2\otimes|0\rangle_b = |0\rangle_2\otimes|-i\alpha_0\rangle_b,
\end{equation}
so that we have swapped the steady state coherent amplitude in the optical cavity into the memory, with a $\pi/2$ phase change. 
We thus find that at the end of the first step of the protocol we have prepared the memory mode $b$ in the coherent state $|\beta\rangle_b$ where $\beta=-2\epsilon/\kappa_2$. At this point in time we turn off the driving field on cavity-2 allowing it to relax back to the vacuum state.  This completes the first step of the protocol.

\section*{Appendix B: Characterization of the effective beam splitter in MZ and HOM interferometers in semiclassical regime}

The light detected in the reflection/transmission port of the effective beam splitter comprises of two parts as shown in figure \ref{fig6}: (1) the field that bounces off the cavity and directly moves from the source to the detector without entering the cavity (dashed blue line in figure \ref{fig6}) and (2) the field which is detected from within the cavity (solid blue line in figure \ref{fig6}). To characterize this effective beam splitter having this in mind, we use the cavity beam splitter in two model system interferometers: a classical MZ interferometer and a quantum HOM interferometer. We wish to study the visibility of the interference pattern to obtain some intuition as to how this effective beam splitter can be compared to a conventional beam splitter.

      \subsection{Characterization of the beam splitter in a Mach-Zender interferometer}
We inject a single photon with an exponentially decaying shape into the interferometer already described in figure \ref{fig2}. The visibility of the interference pattern is given by the relation
\begin{equation}
v =\frac{P_u^{{\rm max}}-P_u^{{\rm min}}}{P_u^{{\rm max}}+P_u^{{\rm min}}},
\end{equation}
where 
\begin{equation}
P_u =\int_0^{\infty} \langle a^\dagger_{1,{\rm out}}a_{1,{\rm out}}\rangle_t dt,
\end{equation}
is the probability to detect a single photon at any time in the upper detector $D_1$. The input state incident on the effective beam splitter (second beam splitter shown in figure \ref{fig2}) after passing the conventional 50/50 beam splitter (first beam splitter shown in figure \ref{fig2}) is 
\begin{equation}
\vert \psi (0)\rangle=\frac{1}{\sqrt{2}}(e^{i\phi}\vert 1_{a_1},0_{a_2}\rangle+\vert 0_{a_1},1_{a_2}\rangle).
\end{equation}
Therefore, $P_u$ becomes 
  
   \begin{equation}
  P_u =\dfrac{4\kappa g\left(4g^2-\kappa(\kappa+\gamma)\right)\sin (\phi)}{\left(4g^2+\kappa^2 \right)\left(4g^2+(\kappa+\gamma)^2\right)} +\dfrac{1}{2}.
  \label{Pu}
   \end{equation}
    \begin{figure}[htbp]
   \centering
   \includegraphics[width=0.9\textwidth]{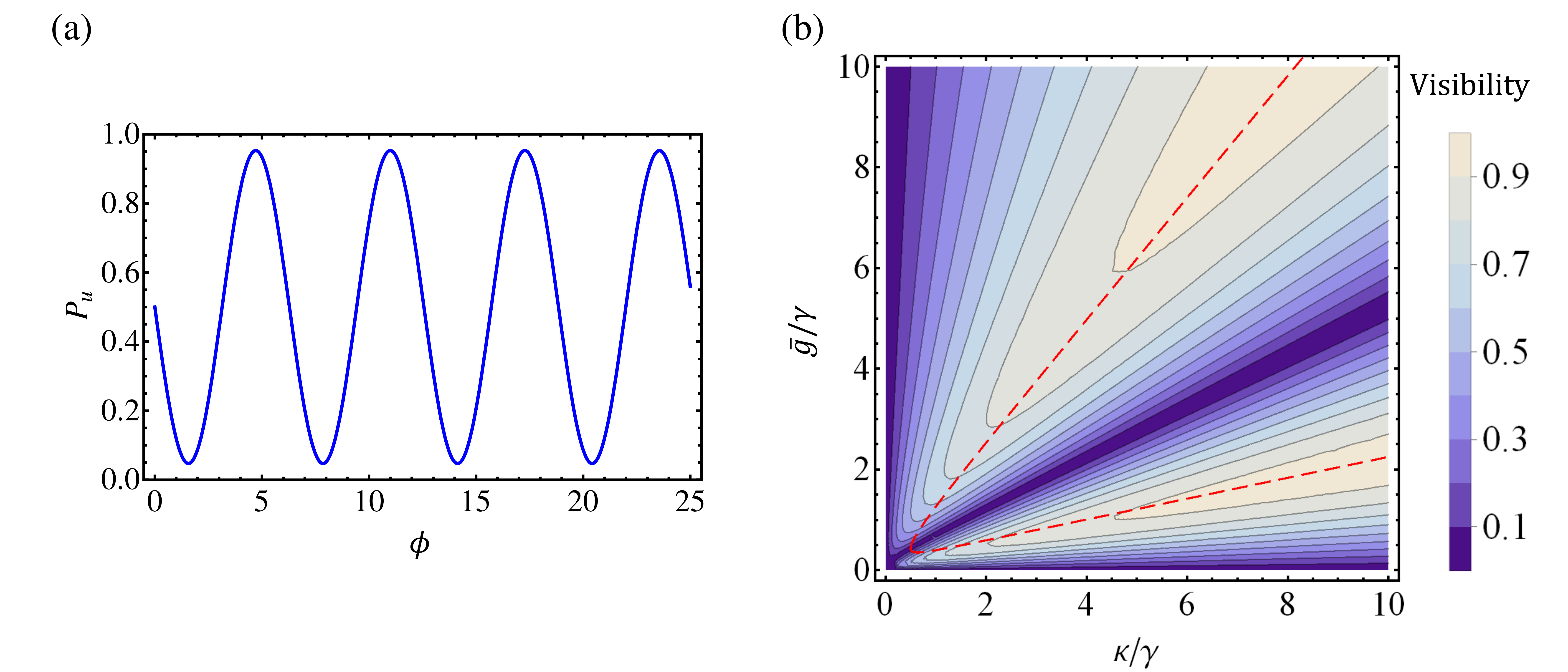} 
  \caption{(a) detection probability at detector $D_1$ versus phase shift. This figure shows the interference pattern for $\gamma=1$, $\kappa=5$ and $\bar{g}=1.2$ which results in a beam splitter reflectivity of $R=0.5$ and interference visibility of $v_{MZ}\simeq0.91$. (b) MZ visibility for different regimes of $\kappa$ and $\bar{g}$ in units of $\gamma$. The red dashed line shows the parameters for which the transmission of the effective beam splitter $T=0.5$.}
   \label{appendix1}
\end{figure}
This relation gives the typical interference pattern for a MZ interferometer shown in figure \ref{appendix1}(a). The interference visibility is given by
\begin{equation}
 v=\vert \dfrac{8\kappa g\left(4g^2-\kappa(\kappa+\gamma)\right)}{\left(4g^2+\kappa^2 \right)\left(4g^2+(\kappa+\gamma)^2\right)}\vert.
\end{equation}
Figure \ref{appendix1}(b) shows the MZ visibility for different values of $\kappa/\gamma$ and $\bar{g}/\gamma$. 

For a MZ interferometer in which the first beam splitter is a conventional 50/50 beam splitter and the second beam splitter is a conventional beam splitter with reflectivity/transmissivity $\mathcal{R}/\mathcal{T}$, the visibility of the interference is $2\sqrt{\mathcal{R}\mathcal{T}}$. For $\mathcal{R}=0.5$, the visibility is one. The red dashed line in figure \ref{appendix1}(b) shows the parameter regime where according to figure \ref{fig1}, the transmission is $0.5$. In this case, to compare the cavity beam splitter with a conventional 50/50 beam splitter, we need to achieve a visibility as close as possible to 1. This figure shows that to achieve a visibility greater than $0.9$, in the case of $T=0.5$, one needs to work in regimes where $\kappa/\gamma\gg1$. However, one can see that for values of $T$ other than $T=0.5$, the overlap of the corresponding transmission contour given in figure (\ref{fig1}) with the expected visibility value can be achieved in regimes where $\kappa/\gamma\simeq1$ or $\kappa/\gamma<1$. 

In the next section, we use the effective beam splitter in a HOM interferometer which demonstrates a fully quantum phenomena.

       \subsection{Characterization of the beam splitter in a Hong-Ou-Mandel interferometer}

The scheme for a HOM interferometer implemented with the cavity beam splitter is shown in figure (\ref{fig6}). We send one photon into each of the cavities $a_1$ and $a_2$. The single photons are specified by the same amplitude function but one of the input photons is time shifted with respect to the other
\begin{eqnarray}
\xi(t)=\sqrt{\gamma} e^{-\frac{1}{2}\gamma t},\nonumber \\
\eta(t)=\sqrt{\gamma} e^{-\frac{1}{2}\gamma (t-\tau)}.
\end{eqnarray}
The joint photon counting probability is given by equation (\ref{G2}). For the initial state $\vert\psi(0)\rangle=\vert1_{a_1,\xi}1_{a_2,\eta}\rangle$ this joint detection probability becomes~\cite{BasiriOpx}
\begin{eqnarray}
G^{(2)}(\delta \tau)=\dfrac{e^{-\dfrac{3}{2}\delta \tau (\kappa+\gamma)}}{A}\left(Be^{\dfrac{3}{2}\delta \tau (\kappa+\gamma)}+Ce^{-\dfrac{1}{2}\delta \tau (3\kappa+\gamma)}
+De^{\dfrac{1}{2}\delta \tau (\kappa+3\gamma)}+Ee^{\delta \tau (\kappa+\gamma)}\right),
\label{G2_sol}
\end{eqnarray}
where
\begin{equation}
A=(4g^2 + \kappa^2)^2\bigg(16 g^4 +(\gamma^2 -\kappa^2)^2+8 g^2 (\gamma^2 + \kappa^2)\bigg)^2,\nonumber
\end{equation}
\begin{eqnarray}
B&=& (4g^2 + (\gamma - \kappa)^2)^2 \bigg(256g^8 + \kappa^4 (\gamma+ \kappa)^4
+8g^2 (\gamma^2 -2 \kappa^2) (16g^4 + \kappa^2 (\gamma+ \kappa)^2)\nonumber \\
&&+16g^4(\gamma^4 + 2\gamma^2 \kappa^2 +20 \gamma\kappa^3 + 22\kappa^4)\bigg),\nonumber
\end{eqnarray} 
\begin{equation} 
C=-32g^2\kappa^2(4 g^2 +\gamma^2 -\kappa^2)^2(4g^2 + \kappa^2)^2,\nonumber
\end{equation}
\begin{equation}
D = -32 g^2 \gamma^2 \kappa^2 F^2,\nonumber
\end{equation}
\begin{equation}
E = -64g^2\gamma\kappa^2(4g^2 +\gamma^2 -\kappa^2)(4g^2 +\kappa^2)F,\nonumber
\end{equation}
and
\begin{eqnarray}
F=\kappa(-12g^2 - \gamma^2 + \kappa^2)\cos(g\delta\tau)
+ 2 g (4 g^2 + \gamma^2 - 3 \kappa^2) \sin(g \delta\tau).\nonumber
\end{eqnarray}
Figure \ref{Ch3_fig6_HOMinterference}(a) shows the HOM interference pattern for some arbitrary parameters $\kappa,$ and $g$. For $\tau=0$, where input photons are indistinguishable, quantum interference results in photon bunching and we see the HOM dip. The visibility is defined as
\begin{equation}
v_{{\rm HOM}}=\dfrac{G^{(2)}(\delta\tau\rightarrow\infty)-G^{(2)}(0)}{G^{(2)}(\delta\tau\rightarrow\infty)+G^{(2)}(0)}.
\end{equation}
Figure \ref{Ch3_fig6_HOMinterference}(b) shows HOM visibility for different values of $\kappa/\gamma$ and $\bar{g}/\gamma$. We see that compared to what we had in the case of a MZ interferometer, HOM visibility is more sensitive to changes in $\bar{g}$ and $\kappa$. Moreover, to work in regime where $R=0.5$ and visibility$>0.9$, we need a larger $\kappa/\gamma$ compared to those needed in MZ interferometer case. We also need to work in stronger coupling regimes. The red dashed line shows the parameters for which the transmission of the effective beam splitter $T=0.5$.
\begin{figure}[htbp]
   \centering
   \includegraphics[width=0.9\textwidth]{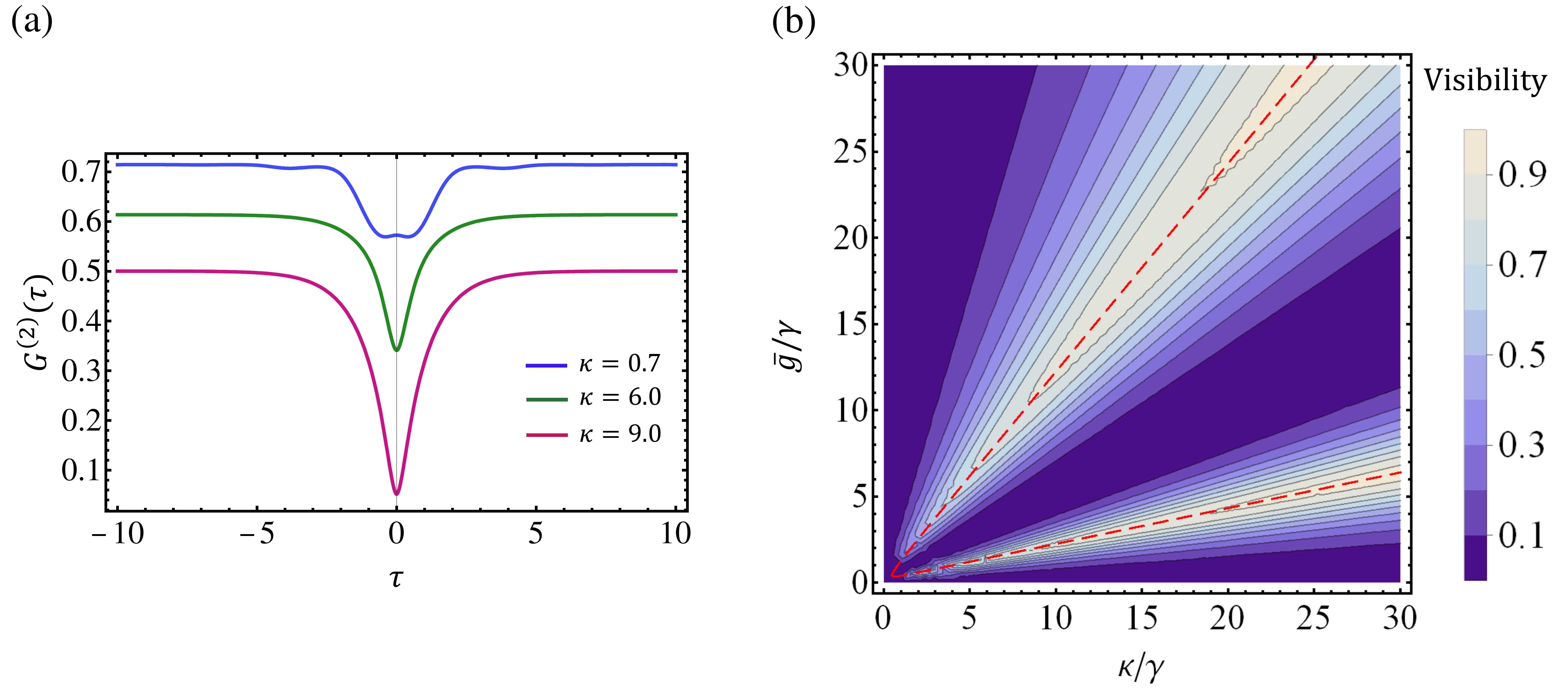} 
  \caption{(a) HOM dips for $\gamma=1$, $\bar{g}=2$ and the shown values of $\kappa$. According to figure (\ref{fig1}), the optical parameters $\bar{g}$, $\kappa$ and $\gamma$ given for the red curve illustrate an effective beam splitter having the reflectivity of $R=0.5$. The interference visibility of this curve is 0.81. (b) HOM visibility presented for different regimes of beam splitter parameters $\kappa$ and $\bar{g}$ in units of $\gamma$.}
   \label{Ch3_fig6_HOMinterference}
\end{figure}

\acknowledgements
Authors wish to thank Ian Walmesly and Ben Baragiola for useful discussions. We acknowledge the support of the Australian Research Council Centre of Excellence for Engineered Quantum Systems, CE110001013. SBE also acknowledges The University of Queensland international scholarship. JC was supported in part by National Science Foundation Grant No. PHY-1212445 and by Office of Naval Research Grant No. N00014-11-1-0082.


\begin{references}
\bibitem{warren1993coherent}Warren W S, Rabitz H and Dahleh M 1993 {\it Science} {\bf 259} 1581
\bibitem{rabitz2000whither}Rabitz H, de Vivie-Riedle R, Motzkus M and Kompa K 2000 {\it Science} {\bf 288} 824
\bibitem{palomaki2013entangling}Palomaki T A, Teufel J D, Simmonds R W and Lehnert K W 2013 {\it Science} {\bf 342} 710
\bibitem{Kolkowitz}Kolkowitz S, Jayich A C B, Unterreithmeier Q P, Bennett S D, Rabl P, Harris J G E and Lukin M D 2012 {\it Science} {\bf 335} 1603
\bibitem{Gavartin}Gavartin E, Verlot P and Kippenberg T J 2012 {\it Nature Nanotech.} {\bf 7} 509
\bibitem{Maiwald}Maiwald R, Leibfried D, Britton J, Bergquist J C, Leuchs G and Wineland D J 2009 {\it Nature Phys.} {\bf 5} 551
\bibitem{BasiriOpx}Basiri-Esfahani S, Myers C R, Armin A, Combes J, Milburn G J 2015 {\it Opt. Express} {\bf 23} 16008

\bibitem{jeong2014coarsening}Jeong H, Lim Y and Kim M S 2014 {\it Phys. Rev. Lett.} {\bf 112} 010402
\bibitem{bennett2000quantum}Bennett C H and DiVincenzo D P 2000 {\it Nature} {\bf 404} 247
\bibitem{knill2001scheme}Knill E, Laflamme R and Milburn G J 2001 {\it Nature} {\bf 409} 46
\bibitem{BowenMilburn}Bowen W P and Milburn G J 2015 {\it Quantum Optomehanics} (CRC Press)
\bibitem{Wollman}Wollman E E, Lei C U, Weinstein A J, Suh J, Kronwald A, Marquardt F, Clerk A A, Schwab K C 2015 {\it Science} {\bf 349} 952 
\bibitem{Pirkkalainen}Pirkkalainen J M, Damskagg E, Brandt M, Massel F and Sillanpaa M A 2015 {\it Phys. Rev. Lett.} {\bf 115} 243601
\bibitem{Lecocq}Lecocq F, Clark J B, Simmonds R W, Aumentado J and Teufel J D 2015 {\it Phys. Rev. X} {\bf 5} 041037 
\bibitem{Galland}Galland C, Sangouard N, Piro N, Gisin N and Kippenberg T J 2014 {Phys. Rev. Lett.} {\bf 112} 143602 
\bibitem{Vanner}Vanner M R, Aspelmeyer M and Kim M S 2013 {Phys. Rev. Lett.} {\bf 110} 010504
\bibitem{NJP-single-photon}Basiri-Esfahani S, Akram U and Milburn G J 2012 {\it New J. Phys.} {\bf 14} 085017 
\bibitem{Borkje}Borkje K 2014 {\it Phys. Rev. A} {\bf 90} 023806  
\bibitem{Akram}Akram U, Bowen W P and Milburn G J 2013 {\it New J. Phys.} {\bf 15} 093007 
\bibitem{Setkaski}Sekatski P, Aspelmeye M and Sangouard N 2014 {\it Phys. Rev. Lett.} {\bf 112} 080502 
\bibitem{Agarwal}Agarwal G S and Huang S 2012 {\it Phys. Rev. A} {\bf 85} 021801 
\bibitem{Stannigel}Stannigel K, Rabl P, Sorensen A S, Lukin M D and Zoller P 2011 {\it Phys. Rev. A} {\bf 84} 042341 
\bibitem{Komar}Komar P, Bennett S D, Stannigel K, Habraken S J M, Rabl P, Zoller P and Lukin M D 2013 {\it Phys. Rev. A} {\bf 87} 013839 
\bibitem{Milburn-singlePhot}Milburn G J and Basiri-Esfahani S 2015 {\it Proc. R. Soc. A.} {\bf 471} 20150280 
\bibitem{Scarani}Scarani V, Bechmann-Pasquinucci H, Cerf N J, Dusek M, Lutkenhaus N N, Peev M 2009 {\it Rev. Mod. Phys.} {\bf 81} 1301
\bibitem{langford2011efficient}Langford N K, Ramelow S, Prevedel R, Munro W J, Milburn G J and Zeilinger A 2011 {\it Nature} {\bf 478} 360 
\bibitem{Collins}Collins M J, Xiong C, Rey I H, Vo T D, He J, Shahnia S, Reardon C, Krauss T F, Steel M J, Clark A S and Eggleton B J 2013 {\it Nature Commun.} {\bf 4} 2582 
\bibitem{Buckley}Buckley S, Rivoire K, Vuckovic J 2012 {\it Rep. Prog. Phys.} {\bf 75} 126503 
\bibitem{Riedinger}Riedinger R, Hong S, Norte RA, Slater JA, Shang J, Krause AG, Anant V, Aspelmeyer M and Gr\" oblacher S 2016 {\it Nature} {\bf 530} 313 

\bibitem{Comandar}Comandar L C, Fr\" ohlich B, Dynes J F, Sharpe A W, Lucamarini M, Yuan Z L, Penty R V, Shields A J 2015 {\it J. Appl. Phys.} {\bf 117} 083109
\bibitem{Ralph2006}Rohde P P and Ralph T C 2006 {\it J. Mod. Opt.} {\bf 53} 1589 

\bibitem{HOM}Hong C K, Ou Z Y and Mandel L 1987 {\it Phys. Rev. Lett.} {\bf 59}, 2044
\bibitem{boson-sampling1}Broome M A, Fedrizzi A, Rahimi-Keshari S, Dove J, Aaronson S, Ralph T C and White A G 2013 {\it Science} {\bf 339}, 794 
\bibitem{boson-sampling2}Spring J B, Metcalf B J, Humphreys P C, Kolthammer W S, Jin X M, Barbieri M, Datta A, Thomas-Peter N, Langford N K, Kundys D, Gates J C 2013 {\it Science} {\bf 339}, 798 

\bibitem{Milburn-RS}Milburn G J 2012 {\it Proc. Roy Soc. A} {\bf 370}, 4469

\bibitem{Yuan}Yuan M, Singh V, Blanter Y M, Steele G A 2015 {\it Nature commun.} {\bf 6} 8491 
\bibitem{Painter}Eichenfield M, Chan J, Camacho R M, Vahala K J and Painter O 2009 {\it Nature} {\bf 462}, 78 
\bibitem{Chang}Chang D E, Safavi-Naeini A H, Hafezi M, Painter O 2011 {\it New J. Phys.}  {\bf 13}, 023003 
\bibitem{stock2011high}Stock E, Unrau W, Lochmann A, T\" offlinger J A, \"Ozt\" urk M, Toropov A I, Bakarov A K, Haisler V A, Bimberg D 2011 {\it Semiconductor Sci. Technol.} {\bf 26} 014003
\bibitem{buckley2012engineered}Buckley S, Rivoire K and Vuckovic J 2012 {\it Rep. Prog. Phys.} {\bf 75} 126503


\bibitem{Marquardt2012}Ludwig M, Safavi-Naeini A H, Painter O and Marquardt F 2012 {\it Phys. Rev. Lett.} {\bf 109}, 063601 
\bibitem{Nunn2008}Nunn J, Reim K, Lee K C, Lorenz V O, Sussman B J, Walmsley I A, Jaksch D 2008 {\it Phys. Rev. Lett.} {\bf 101} 260502 
\bibitem{HMW-GJM}Wiseman H M and Milburn G J 1994 {\it Phys. Rev. A} {\bf 49} 4110 
\bibitem{milburn2008coherent}Milburn G J 2008 {\it Eur. Phys. J. Special Topics} {\bf 159} 113
\bibitem{Chan-cooling1}Chan J, Alegre T M, Safavi-Naeini A H, Hill J T, Krause A, Gr\" oblacher S, Aspelmeyer M and Painter O 2011 {\it Nature} {\bf 478}, 89 
\bibitem{chan2012optimized}Chan J, Safavi-Naeini A H, Hill J T, Meenehan S and Painter O 2012 {\it Appl. Phys. Lett.} {\bf 101} 081115

\bibitem{QObook}Walls D F and Milburn G J 2008 \textit{Quantum Optics} (Springer, Berlin)
\bibitem{Qnoise}Gardiner C W and Zoller P 2004 \textit{Quantum Noise: A Handbook of Markovian and NonMarkovian Quantum Stochastic Methods with Applications to Quantum Optics} (Springer Series in Synergetics, Springer)

\bibitem{fock_josh}Baragiola B Q, Cook R L, Bra\' nczyk A M, Combes J 2012 {\it Phys. Rev. A} {\bf 86} 013811 
\bibitem{fock_zoller}Gheri K M, Ellinger K, Pellizzari T and Zoller P 1998 {\it Fortschr. Phys.} {\bf 46} 401 
\bibitem{BaraCom14} Baragiola B Q and Combes J 2016 {\em in preparation} 


\end{references}
\end{document}